\documentclass[fleqn,usenatbib]{mnras}

\usepackage{graphicx} 
\usepackage{caption,subcaption}

\usepackage[T1]{fontenc}

\DeclareRobustCommand{\VAN}[3]{#2}
\let\VANthebibliography\thebibliography
\def\thebibliography{\DeclareRobustCommand{\VAN}[3]{##3}\VANthebibliography}

\usepackage{graphicx}	% Including figure files
\usepackage{amsmath}	% Advanced maths commands
\usepackage{amssymb}	% Extra maths symbols

\usepackage{newtxtext,newtxmath}

\title[Lithium]{The evolution of Lithium: implications of a universal Spite plateau}

\author[F. Matteucci et al.]{
Francesca Matteucci,$^{1,2,3}$\thanks{E-mail: mn@ras.org.uk (KTS)}
Marta Molero,$^{1}$
David S. Aguado$^{4}$
and Donatella Romano$^{5}$
\\
$^{1}$Department of Physics, University of Trieste, Italy\\
$^{2}$Italian National Institute for Astrophysics (I.N.A.F.) Trieste, Italy\\
$^{3}$Italian National Institute for Nuclear Physics (I.N.F.N.), Trieste, Italy\\
$^{4}$Institute of Astronomy, University of Cambridge, Madingley Road, Cambridge CB3 0HA, UK\\
$^{5}$Italian National Institute for Astrophysics (I.N.A.F.) Bologna, Italy 
}

\date{Accepted XXX. Received YYY; in original form ZZZ}

\pubyear{2021}

\begin{document}
\label{firstpage}
\pagerange{\pageref{firstpage}--\pageref{lastpage}}
\maketitle

\begin{abstract}
The cosmological $^{7}$Li 
 problem consists in explaining why the primordial Li abundance, as predicted by the standard Big Bang nucleosynthesis theory with constraints from WMAP and Planck, is a factor of 3 larger than the Li abundance measured in the stars of the Spite plateau defined by old, warm dwarf stars of the Milky Way halo. Several explanations have been proposed to explain this difference, including various Li depletion processes as well as  non standard Big Bang nucleosynthesis, but the main question remains unanswered. In this paper, we present detailed chemical evolution models for dwarf spheroidal and ultra faint galaxies, compute the galactic evolution of $^{7}$Li abundance in these objects and compare it with observations of similar objects. In our models, Li is mainly produced by novae and cosmic rays and to a minor extent by low and intermediate mass stars. We adopt the yield combination which best fits the Li abundances in the Milky Way stars. It is evident that the observations of dwarf objects define a Spite plateau, identical to that observed in the Milky Way, thus suggesting that the Spite plateau could be a universal feature and its meaning should be discussed. The predictions of our models for dwarf galaxies, are obtained by assuming as Li primordial abundance either the one detected in the atmospheres of the oldest halo stars (Spite plateau; A(Li) $\sim$ 2.2 dex), or the one from cosmological observations (WMAP; A(Li) $\sim$ 2.66 dex). Finally, we discuss the implications of the universality of the Spite plateau results.
 
 %as well as the  formation of the Galactic stellar halo.
\end{abstract}


%%%%%%%%%%%%%%%%%%%%%%%%%%%%%%%%%%%%%%%%%%%%%%%%%%

%%%%%%%%%%%%%%%%% BODY OF PAPER %%%%%%%%%%%%%%%%%%

\section{Introduction}

The study of the evolution of the abundance of $^{7}$Li is very important for cosmology because its primordial abundance, together with the abundances of He and D, allows us to impose constraints on the baryonic density of the Universe ($\Omega_b$). The study of the evolution of $^{7}$Li abundance has started with \cite{Spite1982}, \cite{Spite1986},  \cite{Bonifacio1997} and \cite{Bonifacio2007}, who identified a clear plateau for the Li abundance in dwarf halo stars. \cite{Rebolo1988} showed, for the first time, the relation of the abundance of Li versus metallicity ([Fe/H]) for stars in the solar neighbourhood, and suggested that the upper envelope of such a plot indicates the evolution of the Li abundance during the Galactic lifetime. This is because the large spread in Li abundances, visible especially in disk stars, is the result of different levels of Li destruction in stars in different evolutionary stages. This element, in fact, is easily destroyed by nuclear reactions in stars. Such a spread is not evident, instead, in the Li abundances of halo stars, and the interpretation of this fact has been that Li in halo stars represents the primordial Li abundance, because the plateau value was, at that time,  in agreement with Big Bang nucleosynthesis (BBN) theory. This fact and the small observed spread in halo stars were interpreted as due to the lower convection occurring in low metallicity stars  with consequent lack of Li destruction.
The upper envelope of the data in the Figure 3 of \cite{Rebolo1988} shows a plateau for [Fe/H]$<$ -1.0 dex, followed by a steep rise of the Li abundance in disk stars, reaching the highest value in TTauri stars, where Li is not yet destroyed, and meteorites, and this value is a factor of ten higher than the Spite plateau. An alternative interpretation of this plot is that the primordial Li abundance is that observed in meteorites and the Li in halo stars is the result of its destruction during stellar evolution (see \citealp{Mathews1990}. However, in this case a non-standard Big Bang theory would be required.
Before the WMAP results, the theoretical primordial abundances of all light elements were in agreement with the observed values (e. g. \citealp{Ryan2000}), thus suggesting a value for the baryon to photon ratio ($\eta$) and therefore for $\Omega_b$. The WMAP results (\citealp{Spergel2003}), later confirmed by Planck (\citealp{Coc2014}), suggested directly a precise value for $\eta$ and  therefore a value for the primordial Li abundance. Such a value (A(Li)= 2.66-2.73 dex) is higher by a factor of 3 than the one suggested by the Spite plateau (A(Li)$\sim$ 2.2 dex), thus creating what is called the {\it Cosmological Lithium Problem}: why halo stars show an almost constant Li abundance lower than the primordial one? Before discussing the various proposed solutions to the Li problem, we remind how $^{7}$Li is produced in stars: there is practically only one channel for Li production and in particular there should be a site where the reaction $^{3}He + \alpha \rightarrow ^{7}Be  + \gamma$ can occur, but $^{7}$Be should be rapidly transported by convection in regions of lower temperatures where it decays into $^{7}$Li, and this is known as the \cite{Cameron1971} {\it mechanism}. The proposed stellar sources of $^{7}$Li are: red giant (RG), asymptotic giant branch stars (AGB), novae and supernovae core-collapse. \cite{Dantona1991} computed the Li  evolution in the solar vicinity by means of a detailed chemical evolution model and concluded that RG, AGB and novae should all contribute to Li production, and in particular that only novae, exploding with long time delays (being binary systems of low mass stars), could reproduce the steep rise of the Li abundance at [Fe/H] $>$ -1.0 dex, whereas AGB stars could not and RGs give only a negligible contribution. Later on, other papers included all or part of above mentioned sources in Galactic models (see \citealp{Romano1999, Romano2001, Travaglio2001, Prantzos2017}). The Li production by the neutrino $(\nu)-$process (\citealp{Woosley1990}) was included by \cite{Matteucci1995} as one of the Li sources, but this contribution is also negligible. The $\nu$-process occurs in the shell above the collapsing core in a pre-supernova. The high flux of neutrinos evaporates neutrons and protons from heavy nuclei with the result of producing Li.

Various explanations have been put forward to explain the Spite plateau Li abundance lower than the primordial one. They are: turbulent mixing (\citealp{Richard2005}), gravity waves in stellar interiors (\citealp{Charbonnel2005}), pre-galactic Li processing by massive Population III stars (\citealp{Piau2006}), tachocline mixing induced by rotation (\citealp{Spiegel1992, Piau2008}), mass dependence of Li depletion (\citealp{Melendez2010}), pre-Main Sequence depletion plus accretion (\citealp{fu2015, Molaro2012}) and finally, non-standard Big Bang nucleosynthesis (\citealp{Jedamzik2006, Hisano2009}). What remains difficult to explain is how such processes can deplete Li at the same level in all halo stars.
Moreover, even the Milky Way  Spite-plateau itself has been challenged in the last years, since very metal-poor stars have been found with Li abundances lower than the Spite-plateau (e.g. \citealp{Sbordone2010, Bonifacio2015}), although in \cite{Aguado2019} is presented a star with [Fe/H]$<$-6.0 dex with a Li abundance lying on the Spite-plateau.
This star has been observed in high resolution with UVES at VLT and its inferred Li abundance is A(Li)=$2.02 \pm 0.08$.
These extremely metal-poor stars are very rare and so far only 14 Galactic halo stars with [Fe/H]$<$ -4.5 dex have been observed (\citealt{chris04, fre05, 2012caffau, 2015allende, 2014hansen, kel14, boni15, 2018starkenburg, Agu18II, Agu18I, nord19}) and only 7 of them are unevolved.
More recently, \cite{Gao2020} have claimed the existence of two plateaus in the halo stars, one warm and one cool, with the warm one being at the level of the BBN Li primordial value.

Observations of extragalactic objects have confirmed the existence of the Spite plateau in the globular cluster M54 (\citealp{Mucciarelli2014}), in Sculptor dwarf galaxy (\citealp{Hill2019}) and in $\omega$ Centauri (\citealp{Monaco2010}).  More recently, \cite{Molaro2020} have found the Spite plateau also in Gaia-Enceladus (\citealp{Belokurov2018, Myeong2018, Helmi2018, Molaro2020}), while \cite{Aguado2021} measured the Li abundance in three S2 Stream members, and derived values fully compatible with the plateau. These results show that the behaviour of $^{7}$Li abundance is the same in the Milky Way halo stars, accreted satellites and extragalactic objects. This fact suggests that the Li abundance of the oldest stars is not affected by environmental effects. 
%It is also worth reminding that \cite{Howk2012} measured $^{7}$Li abundance in the ISM of SMC and this is almost equal to the Spite plateau Li value (A(Li)$\sim$ 2.27).\\

In this paper, we present model results relative to the Li abundance for some dwarf spheroidal (dSphs) and ultra faint galaxies (UfDs) and compare them with data. The Li data are taken from a compilation by \cite{Aguado2021} and they show clearly that the Spite plateau exists in all these objects for metallicities lower than [Fe/H]=-1.0 dex. Actually, being all the observed objects metal-poor, the more metal-rich stars with higher Li abundances are likely to be negligible.  The common feature of the Spite plateau among dwarf galaxies and Galactic halo is here discussed in the light of cosmology and galaxy formation and evolution.

The paper is organized as follows: in Section 2 we describe the observational data, in Section 3 the chemical evolution 
models and in Section 4 the results. Finally, in Section 5 we present a discussion and conclusions.

\section{Observational data}
The observed Li abundances we adopted in this paper are the same shown in Figure A1 of \cite{Aguado2021}. These data include stars from the Galactic halo, Sculptor, S2 and Slygr streams, the globular cluster M54, Omega Centauri and Gaia-Enceladus.
From the theoretical point of view we have modeled several dSphs: Sculptor, Fornax, Carina, Draco, Ursa Minor, Sextans and Sagittarius, plus two UfDs: Reticulum II and Bootes I. For all these galaxies we have adopted the metallicity ([Fe/H]) distribution functions  (MDF) from SAGA database.
All these objects are on average metal-poor as the stars in the Galactic halo, and their [Fe/H] extends from -5.0  to -1.0 dex.

\begin{figure}
    \centering 
%\begin{subfigure}{0.30\textwidth}
  \includegraphics[width=\linewidth]{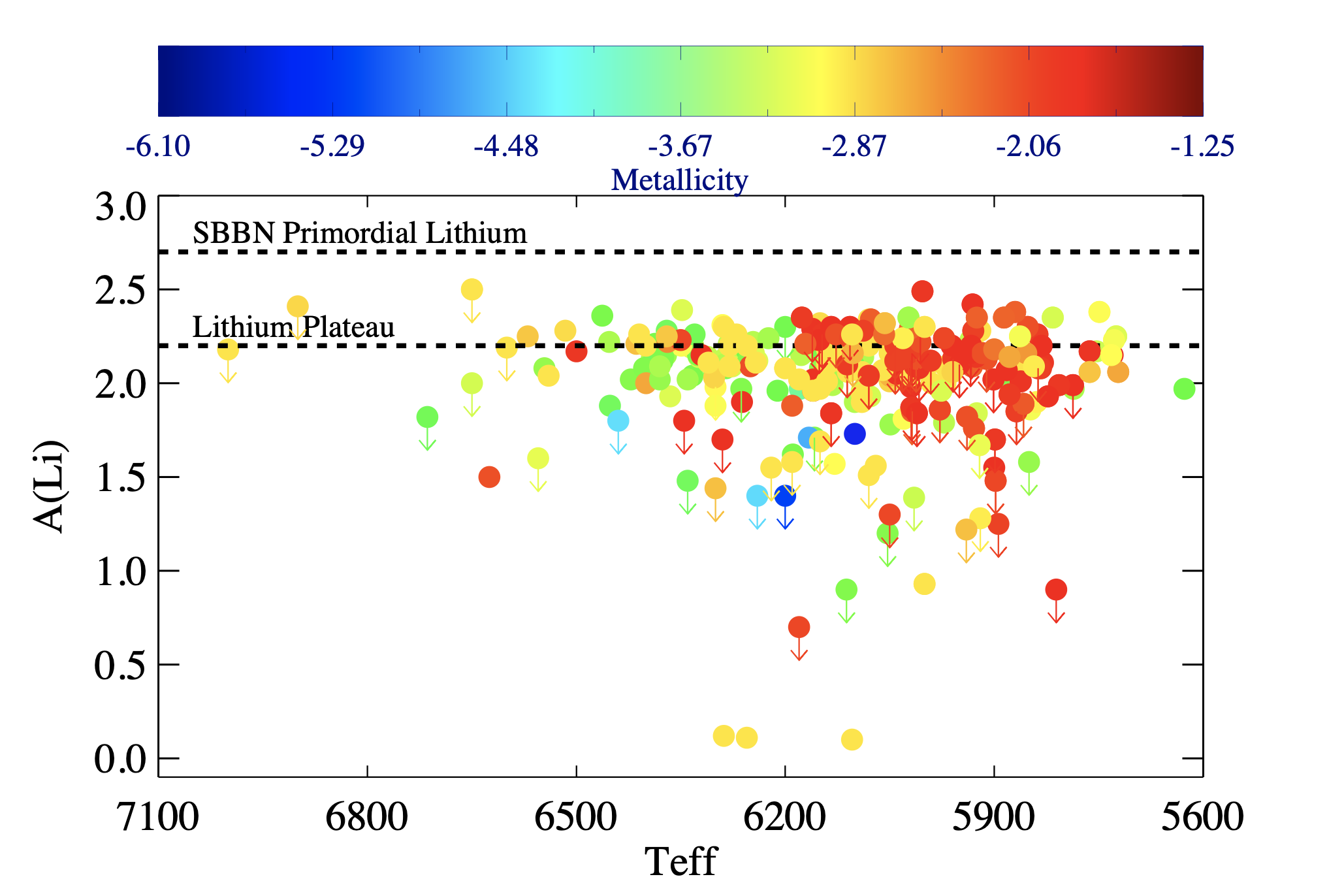}
  \caption{A plot of A(Li) vs. ${T_{\rm eff}}$. Only stars with log(g) $>$3.7 are included. The points are stars from the Galactic halo, S2, Slygr, M54 amd Omega Cen, respectively, taken from the compilation of \citet{Aguado2021}. The colors refer to the stellar metallicity.}
  \label{fig:Lithium_temp}
\end{figure}

In Figure \ref{fig:Lithium_temp} we show  a compilation of  A(Li) values as functions of the effective temperature (${T_{\rm eff}}$) for stars of the Galactic halo, S2, Slygr, M54 and Omega Cen. We include in the plot only stars with log(g)$>$3.7, in order to be sure that no diffusion with lower layers could affect the Li values. In the same plot we show the metallicities of the stars: as one can see, the Li Spite plateau is evident and is neither affected by ${T_{\rm eff}}$ nor metallicity. In the Figure it is also indicated the primordial Li value as suggested by BBN.

\section{Chemical evolution models for dwarf galaxies}

In this paper we present chemical evolution models devised for specific dSphs and UfDs. Both galaxy types are characterized by having suffered a very low star formation rate (SFR) compared to the Milky Way, as indicated by the [$\alpha$/Fe] ratios lower than Galactic halo stars at the same metallicity, except for a small overlapping in the lowest part of the metallicity range. This behaviour of the [$\alpha$/Fe] ratios has been interpreted as due to a very inefficient star formation in these systems (e.g. \citealp{Lanfranchi2003, Salvadori2009, Matteucci2012, Vincenzo2014, Romano2015}). The $\alpha$-ratios in UfDs are even smaller than those in the dSphs and this is because in a low star formation regime, the Fe abundance increases very slowly and when the contribution to Fe from Type Ia SNe occurs, the gas  out of which stars form is still very metal-poor, so we see low [$\alpha$/Fe] ratios at low [Fe/H].\\
The models we have developed here, well reproduce the [$\alpha$/Fe] vs. [Fe/H] plots and the MDFs of each galaxy (Molero et al. submitted).

\subsection{Model prescriptions}

We adopt a model similar to that of \cite{Lanfranchi&Matteucci04} and \cite{Vincenzo2014} to describe the chemical evolution of both UfDs and dSphs. It is assumed that galaxies form by infall of primordial gas falling into the potential well of a diffuse dark matter halo, which is 10 times more massive than the baryonic content. The models are one-zone and assume instantaneous mixing approximation but relax the instantaneous recycling approximation, in order to follow in detail the evolution of the abundances of several light and heavy elements, during $14$ Gyr. For more details about the model prescriptions, we address the reader to \cite{Vincenzo2014}.
The evolution with time of the gas mass in the form of the element $i$, $M_{g,i}(t)$, within the ISM is:
\begin{equation}
\begin{split}
  \dot{M}_{g,i}(t) = & -\psi(t)X_i(t) + (\dot{M}_{g,i})_{inf}-(\dot{M}_{g,i})_{out} + \dot R_{i}(t),
\end{split}
\end{equation}
where $X_i(t)=M_{g,i}(t)/M_{gas}(t)$ is the abundance by mass of the element $i$ at the time $t$ and $M_{gas}(t)$ is the total gas mass of the galaxy.
The terms on the right-hand side of the equation are:

\begin{itemize}
    \item The first term is the rate at which chemical elements are subtracted from the ISM to be included in stars. $\psi(t)$ is the star formation rate (SFR) and is a Schmidt-Kennicutt law with $k=1$, (\citealp{schmidt, kennicutt}):
    \begin{equation}
        \psi(t)=\nu {M_{gas}}^k,
    \end{equation}
    where $\nu$ is the star formation efficiency ($\nu=(0.005-1) Gyr^{-1}$, depending on the galaxy). Here we adopt the star formation histories  (SFHs) as derived by color-magnitude diagrams (CMD) fitting analysis of several authors (\cite{Hernandez2000}; \cite{Dolphin2002}; \cite{deBoer2012}; \cite{fornaxdeBoer2012}; \cite{Brown2014}; \cite{deBoer2015}). The adopted SFH for each galaxy, is characterized in terms of the number ($n$), time of occurrence($t$) and duration ($d$) of the bursts, as reported in Table \ref{tab: SF}. 
    \item The second term is the rate at which chemical elements are accreted through primordial infalling gas: 
    \begin{equation}
        (\dot{M}_{g,i})_{inf}=aX_{i,inf}e^{-t/\tau_{inf}},
    \end{equation}
    where $a$ is a normalization constant constrained to reproduce the present time total gas mass, $X_{i,inf}$ is the abundance of an element $i$ in the infalling gas,  and $\tau_{inf}$ is the infall time-scale ($0.005-3$ Gyr depending on the particular galaxy).
    \item The third term is the rate at which chemical elements are lost through galactic winds. It is assumed to be proportional to the SFR:
    \begin{equation}
        (\dot{M}_{g,i})_{out}=-\omega\psi(t),
    \end{equation}
    where $\omega$ is an adimensional free parameter describing the efficiency of the galactic wind ($\omega=1-12$) and depends on the particular galaxy. 
    \item The last term $R_i(t)$ represents the fraction of matter which is returned by stars into the ISM through stellar winds, supernova explosions and neutron star mergers, in the form of the element $i$. $R_i(t)$ depends on the initial mass function (IMF): here we adopt a \cite{salpeter} IMF for all galaxies with a slope x=1.35 over a mass range of 0.1-100$M_{\odot}$.
    
\end{itemize}

\begin{table}
\centering
\caption{\label{tab: SF} SF histories of dSphs and UfDs. In the $1^{st}$ column it is reported the name of the galaxy and in the $2^{nd}$, $3^{rd}$ and $4^{th}$ column the number, time and duration of the bursts of SF, respectively.}
\begin{tabular}{lccc}
\hline
  Galaxy & $n$ & $t$ $(Gyr)$ & $d$ $(Gyr)$\\
\hline
 Bo\"otes I (BooI) & $1$ & $0$ & $1$\\
 Carina (Car) & $4$ & $0-2-7-9$ & $2-2-2-2$\\
 Draco (Dra) & $1$ & $6$ & $4$\\
 Fornax (For) & $1$ & $0$ & $14$\\
 Reticulum II (RetII) & $1$ & $0$ & $1$\\
 Sagittarius (Sgr) & $2$ & $0-4.5$ & $4-2.5$\\
 Sculptor (Scl) & $1$ & $0$ & $7$\\
 Sextans (Sex) & $1$ & $0$ & $8$\\
 Ursa Minor I (Umi) & $1$ & $0$ & $3$\\
\hline
\end{tabular}%
\end{table}

\subsection{Stellar yields}

For all the elements but $^{7}$Li, the stellar nucleosynthesis prescriptions are as in Model 15 of \cite{Romano2010}. 
%For the yields from SNe Ia (white dwarfs in binary systems) we have adopted the yields of %\cite{Iwamoto1999}. 
For the yields of Li we have considered three sources: i) RG and AGB stars, ii) novae and iii) cosmic rays.
For RG and AGB stars (i.e. low- and intermediate masses $0.8 \le M/M_{\odot} \le$ 8), we adopt the Li yields by \cite{Karakas2010} and \cite{Doherty2014}, \cite{Doherty20142}, that cover also the mass range of super-AGB stars (7-9$M_{\odot}$). However, the Li production from these stars is a minor one. The major contribution to the local abundance of Li in our model is due to nova outbursts, which are strong explosions occurring on the surface of a white dwarf in a binary system, when the less evolved companion (either a main sequence or a giant star) fills its Roche lobe (i.e. \citealp{Starrfield1978}). The explosions do not destroy the white dwarf and each binary system suffers $\sim 10^4$ outbursts on the average during its lifetime (\citealp{Bath1978}). As for the Li production in a single outburst, we choose the amount suggested by \cite{Izzo2015} based on observations of the nova V1369Cen: in particular, the assumed mass ejected by a nova system during a single outburst is $M_{Li} = 2.55 \cdot 10^{-10} M_{\odot}$. It should be noted that this Li detection has not been yet confirmed and that many theoretical uncertainties are present in the Li nova yields (see \citealp{Jose2007}).
For Li production by Galactic cosmic rays we adopt the metallicity-dependent rate of Li production through spallation processes as deduced by \cite{Lemoine1998} from theoretical models. It is worth stressing that by assuming an empirical law (e.g., \citealp{Grisoni2019}) we obtain similar  results.
 We do not include the $\nu$-process by massive stars since it is negligible (it makes $\sim 9\%$ of the meteoritic abundance, as shown by \citealp{Romano2001}).

\begin{figure*}
\begin{center}
    \subfloat{\includegraphics[width=1\columnwidth]{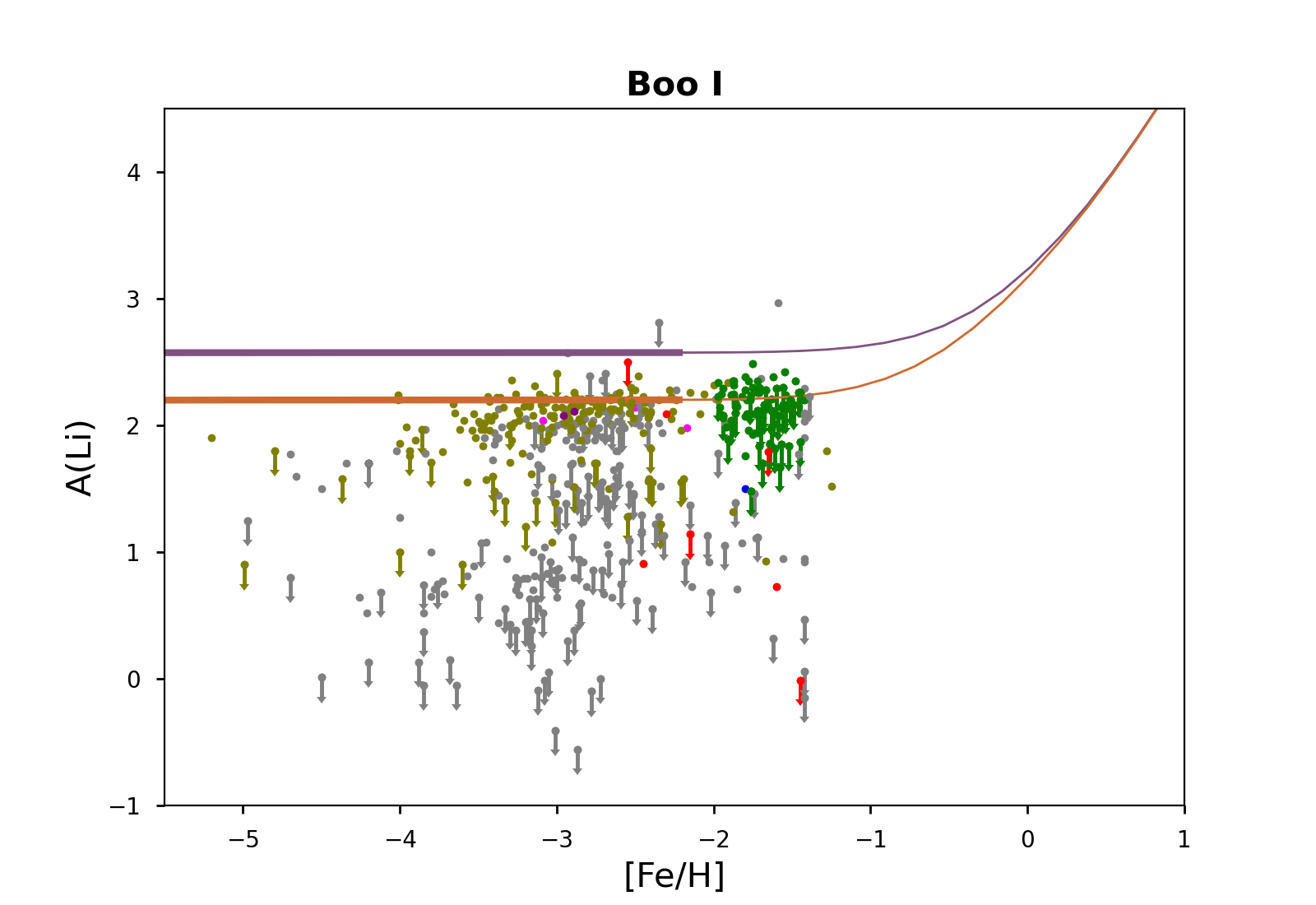}}
    \subfloat{\includegraphics[width=1\columnwidth]{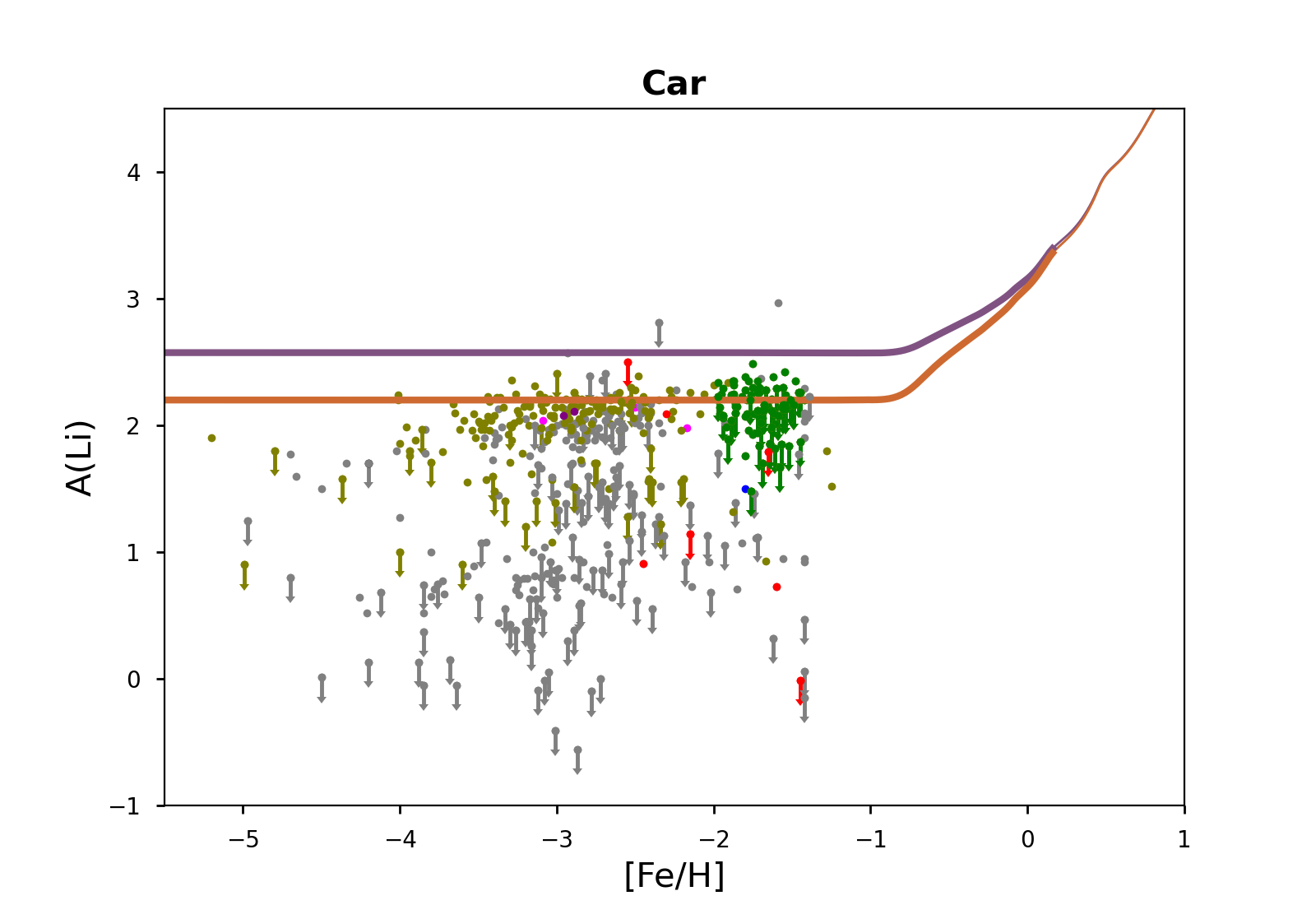}}
    \hfill
    \subfloat{\includegraphics[width=1\columnwidth]{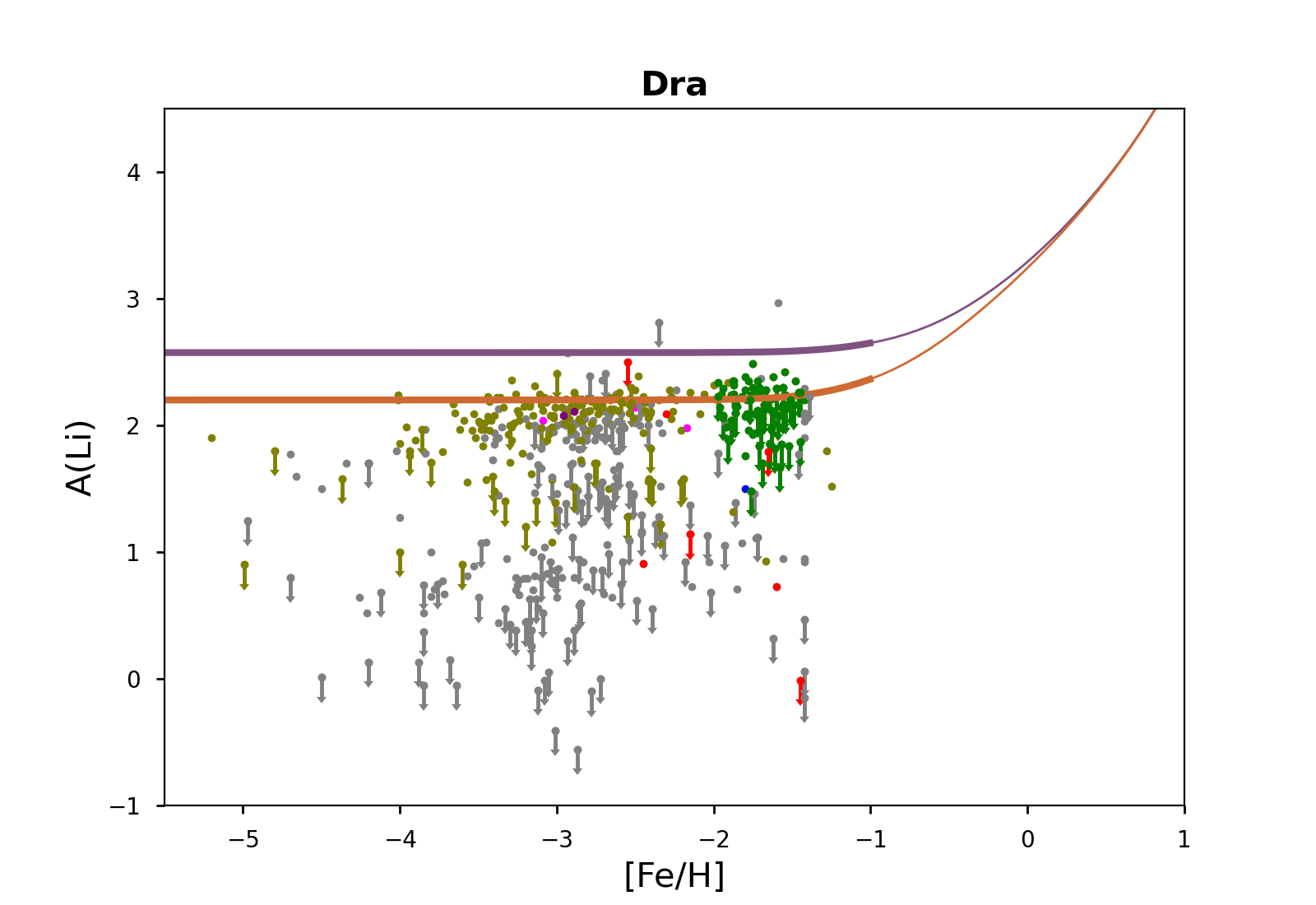}}
    \subfloat{\includegraphics[width=1\columnwidth]{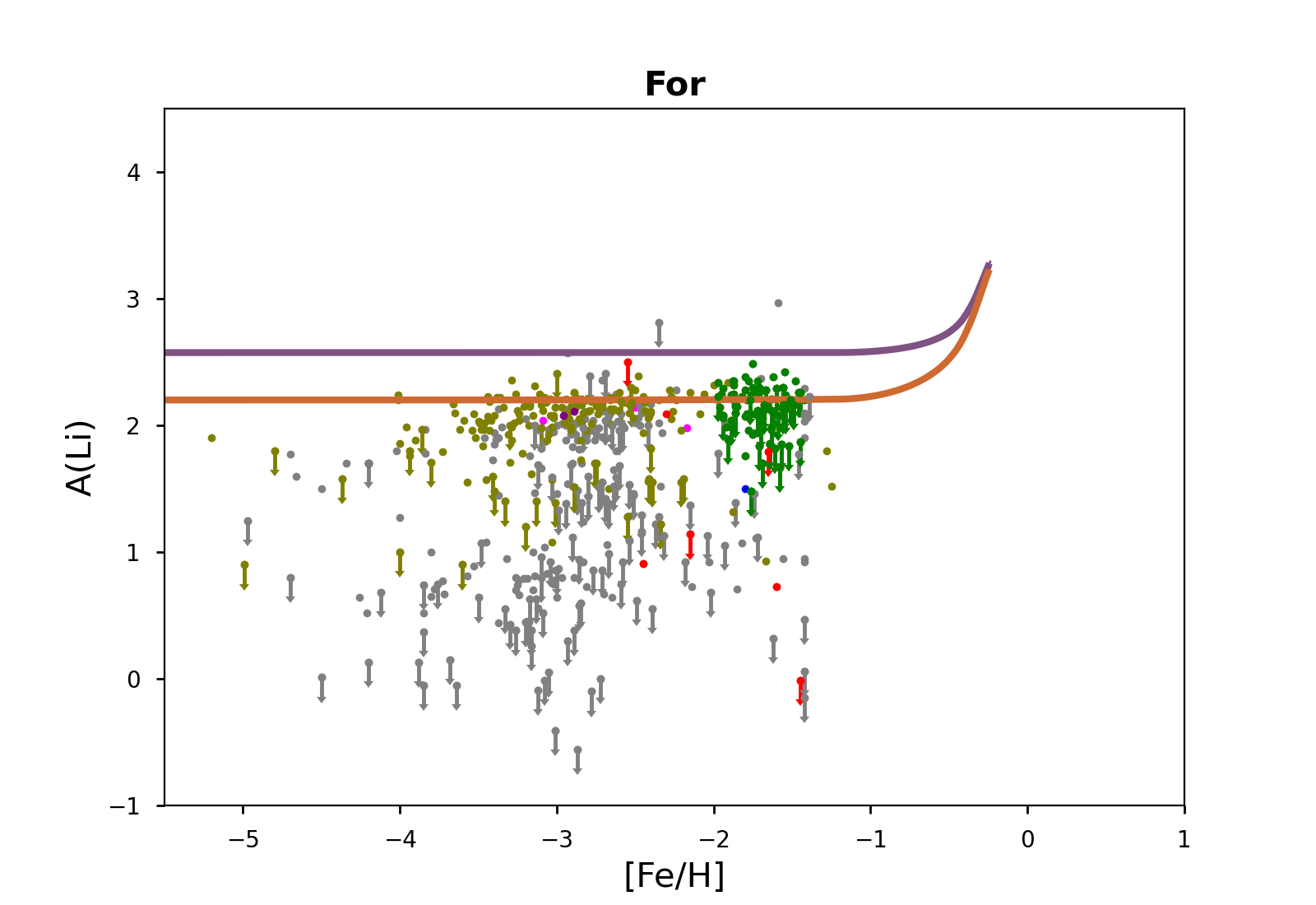}}
    \hfill
    \subfloat{\includegraphics[width=1\columnwidth]{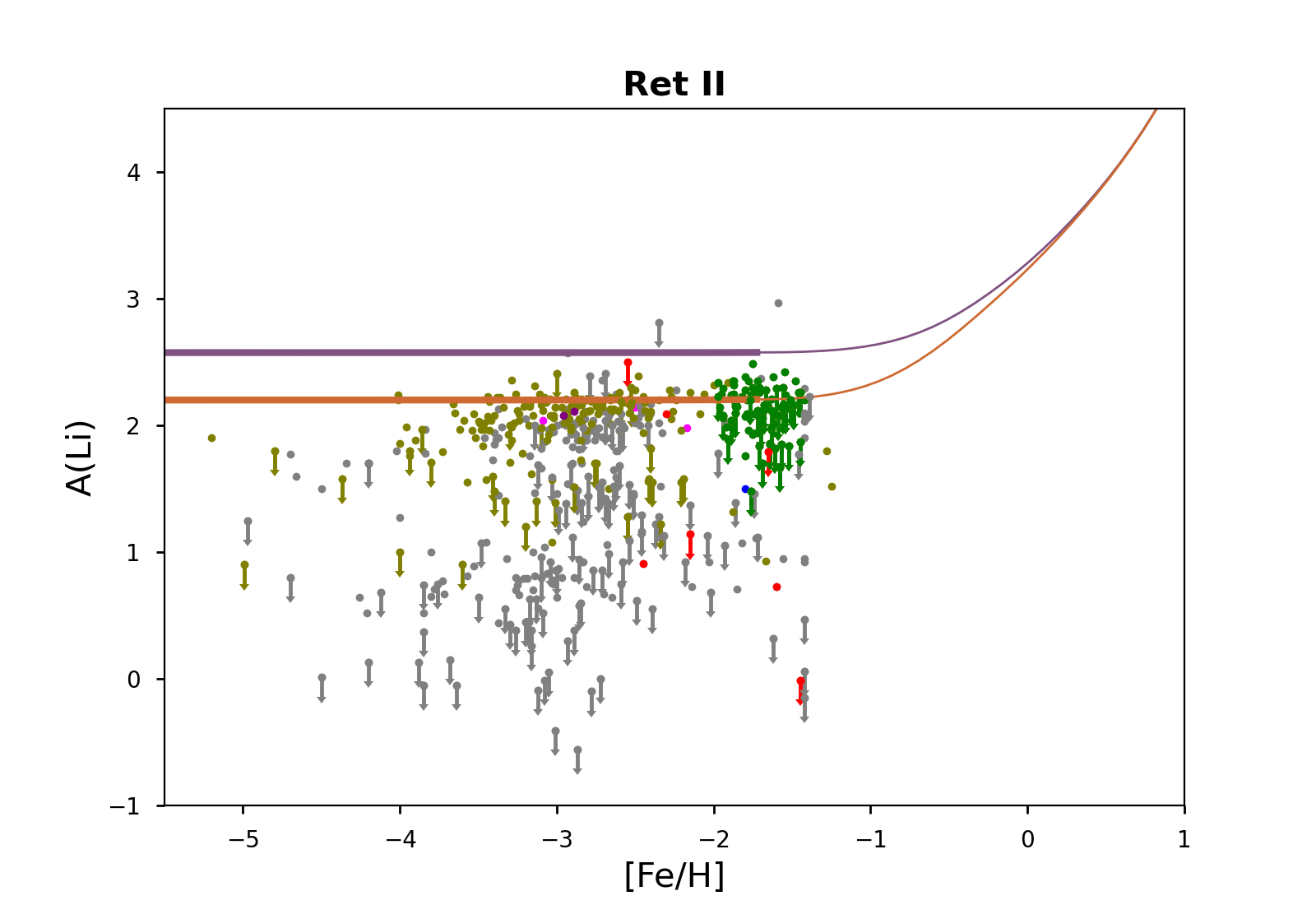}}
    \subfloat{\includegraphics[width=1\columnwidth]{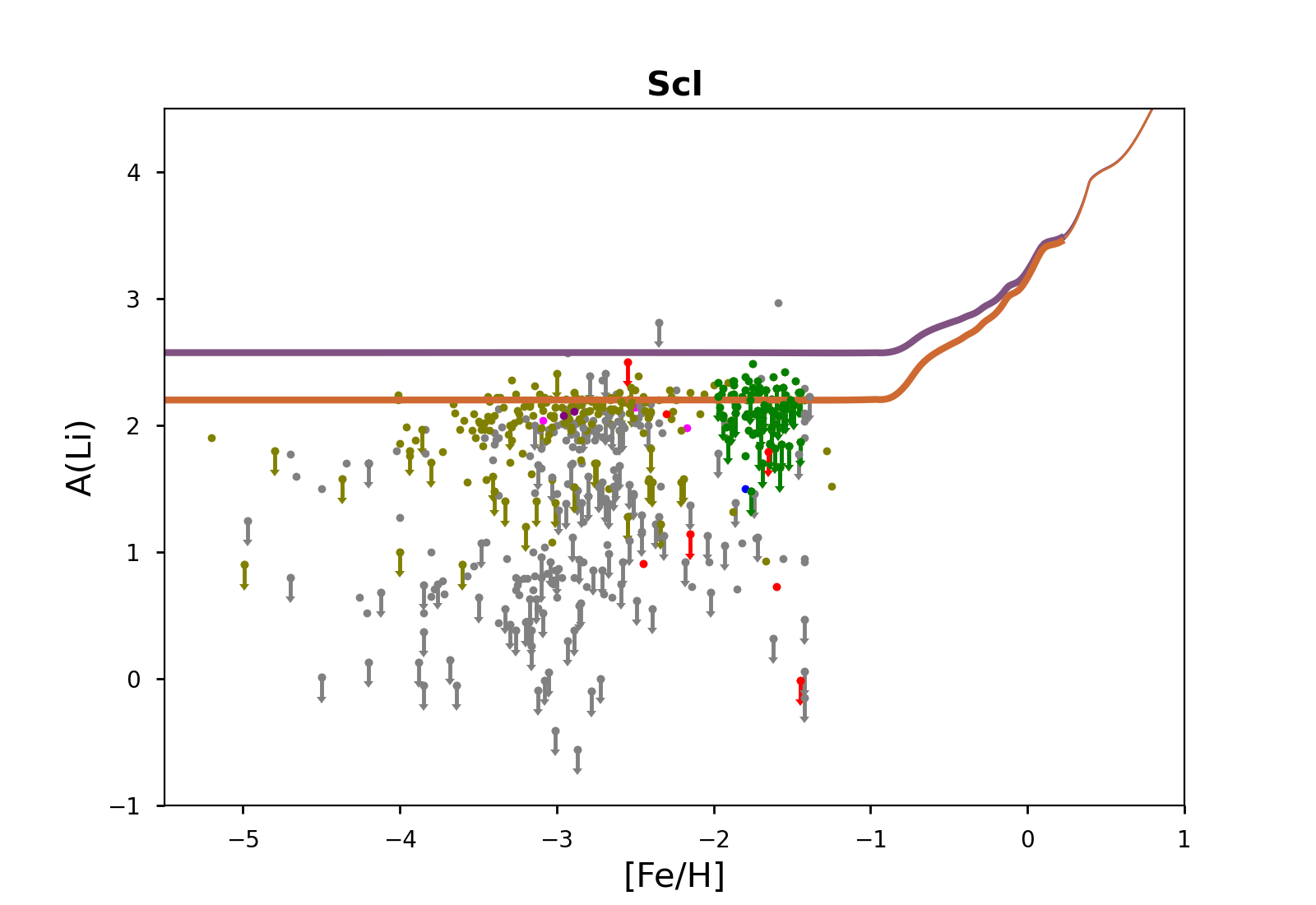}}
\caption{Predicted against observed Lithium abundances versus metallicity for Bootes I, Carina, Draco, Fornax, Reticulum II and Sculptor. Observational data are: A(Li) values of stars from S2 (red dots), from $\omega$ Centauri (green dots), the globular cluster M54 (black dots), Sculptor galaxy (blue dots), the Slygr stream (purple dots), from Gaia-Enceladus (magenta; \citealp{Molaro2020}) and from Galactic halo stars (grey dots). Among these last ones, we highlighted in olive turn off stars (log(g)$\ge3.7$). All data are from \citet{Aguado2021} (see also references therein). The purple lines refer to the BBN primordial Li from WMAP and Planck, while the orange lines refer to a primordial Li with the value of the Spite plateau (2.2 dex). The thick lines, in both colors, refer to the metallicity range of existence of the stars in each galaxy (see also Figure \ref{fig:MDF}), whereas the thin lines refer to a metallicity range where stars are negligible and therefore represent only the abundances in the ISM.}%
\label{fig:Lithium_Both 1}%
\end{center}
\end{figure*}

\begin{figure}
\begin{center}
    \subfloat{\includegraphics[width=1\linewidth]{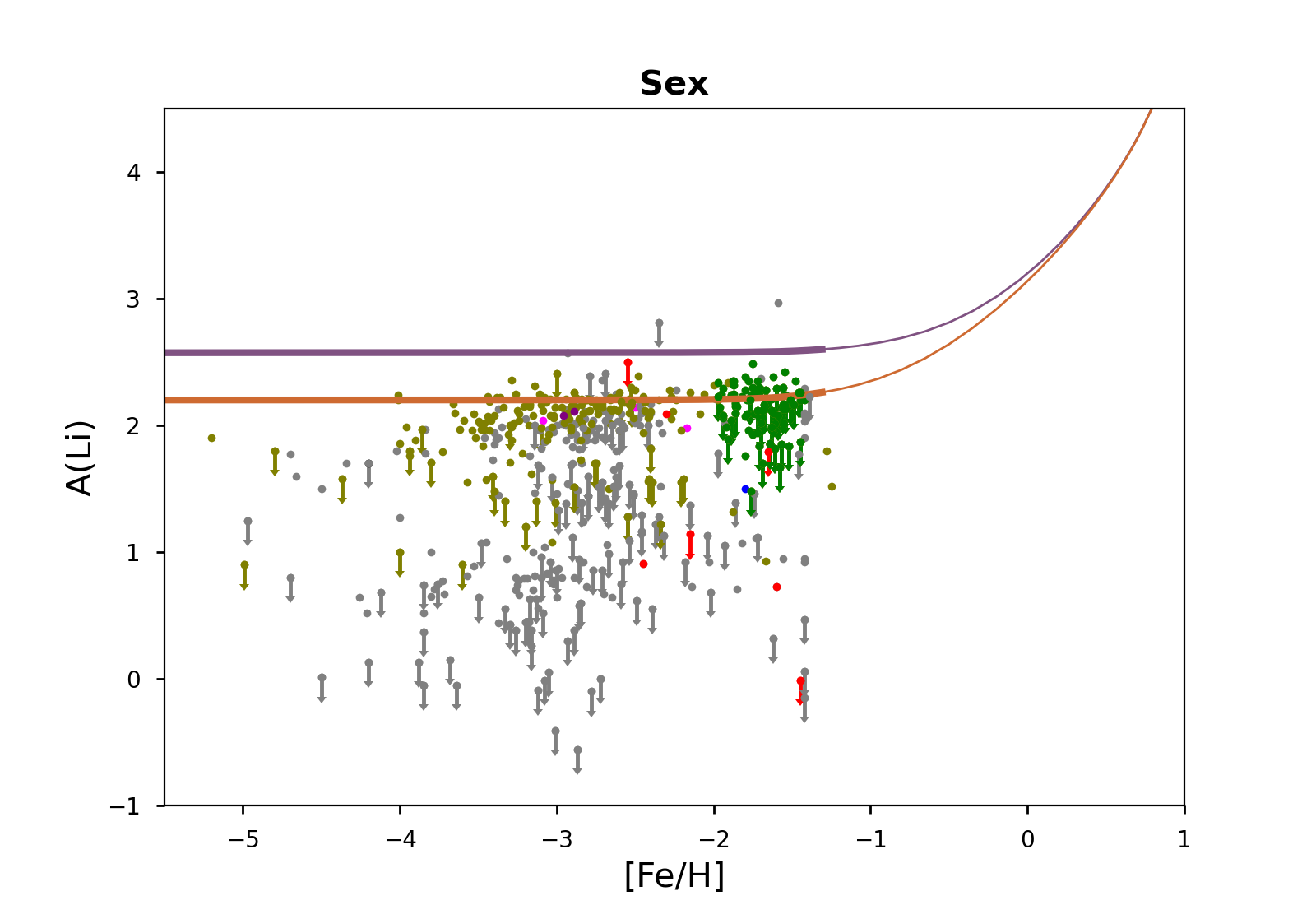}\label{fig:a}}
    \hfill
    \subfloat{\includegraphics[width=1\linewidth]{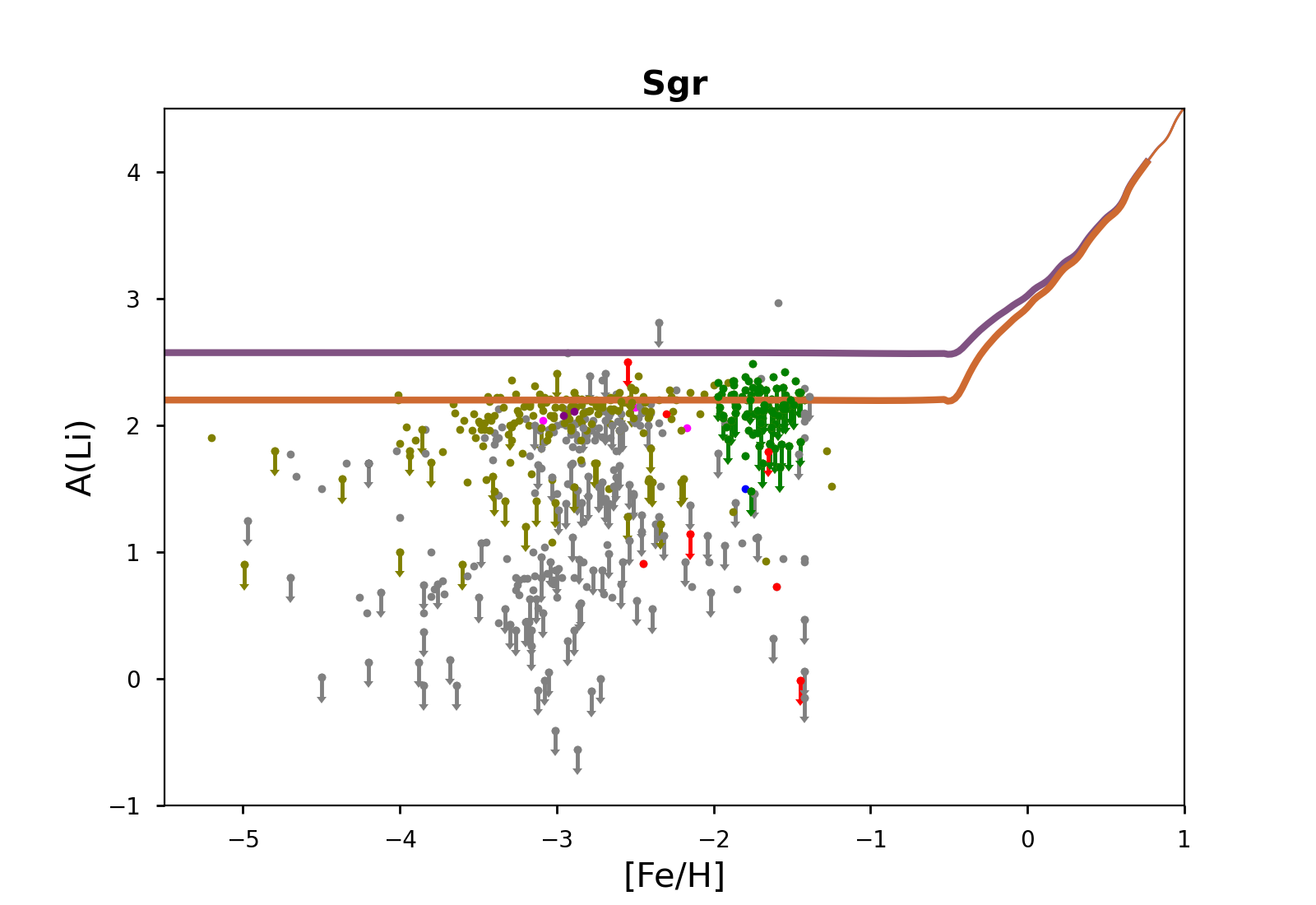}\label{fig:a}}
    \hfill
    \subfloat{\includegraphics[width=1\linewidth]{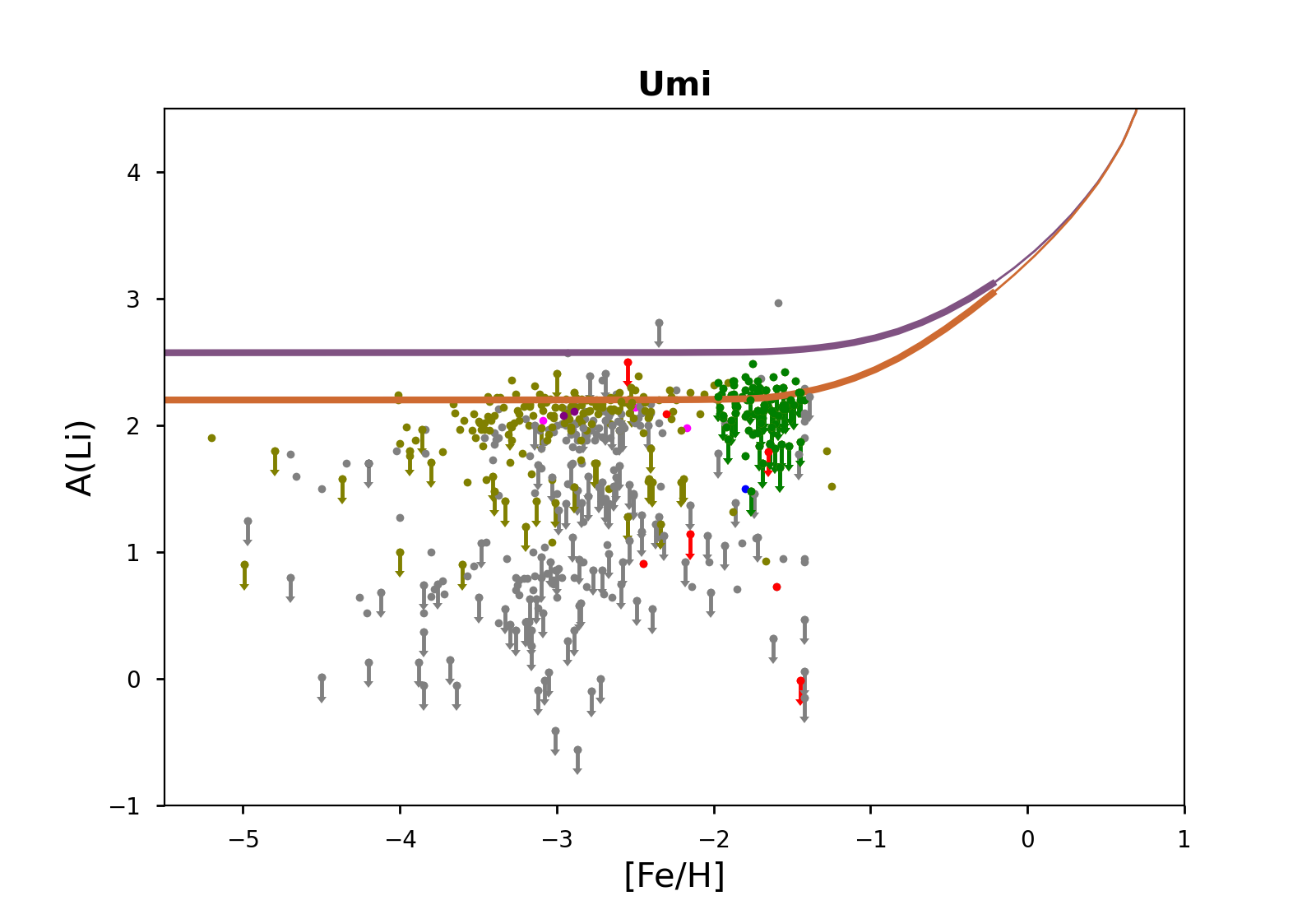}\label{fig:a}}
\caption{Predicted against observed Lithium abundances versus metallicity for Sextan, Sagittarius and Ursa Minor I. Color code and data are as those described in Figure \ref{fig:Lithium_Both 1}.}%
\label{fig: Lithium_Both 2}%
\end{center}
\end{figure}

\begin{figure*}
    \centering 
\begin{subfigure}{0.30\textwidth}
  \includegraphics[width=\linewidth]{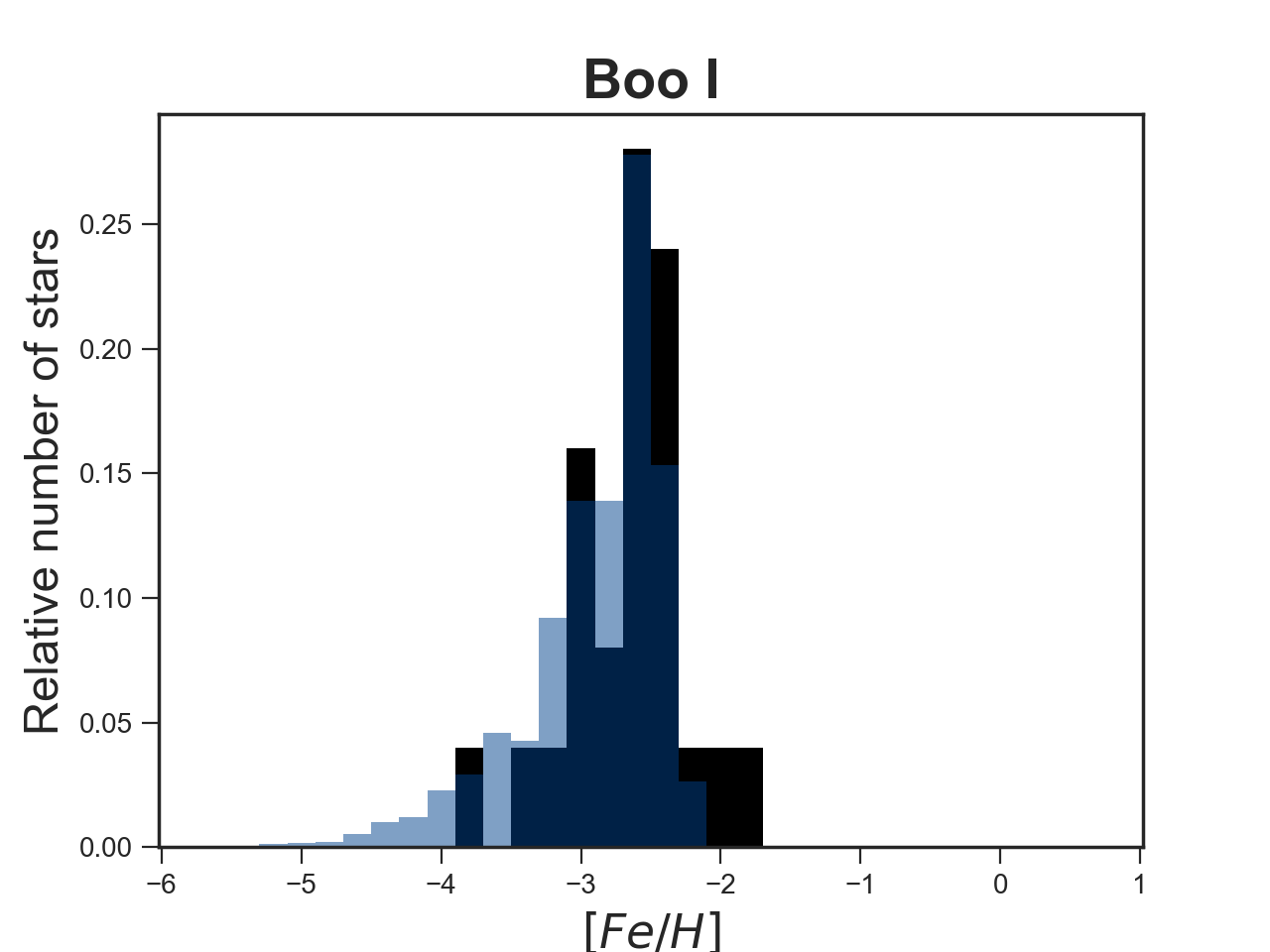}
  \label{fig:1}
\end{subfigure}\hfil 
\begin{subfigure}{0.30\textwidth}
  \includegraphics[width=\linewidth]{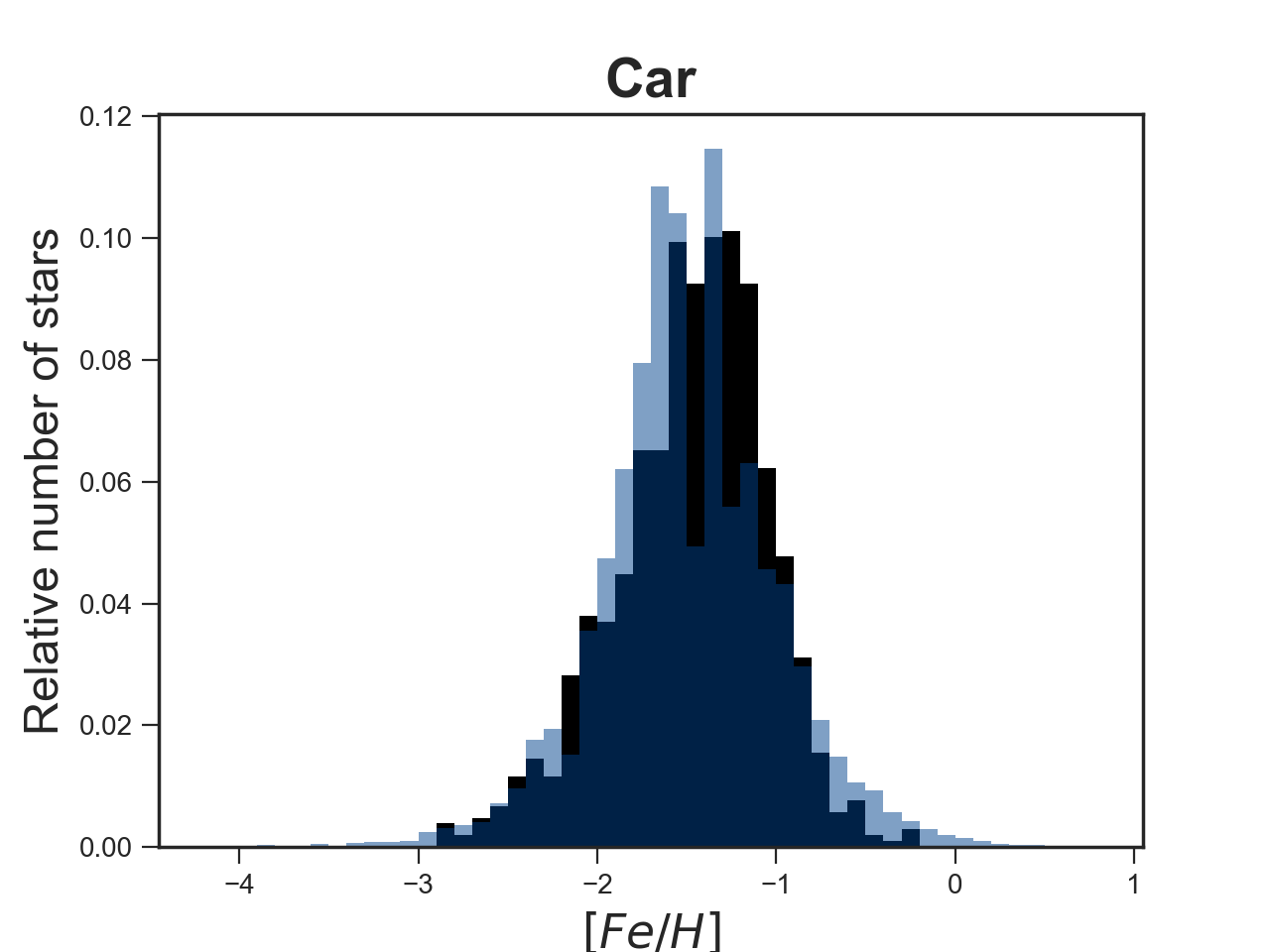}
  \label{fig:2}
\end{subfigure}\hfil 
\begin{subfigure}{0.30\textwidth}
  \includegraphics[width=\linewidth]{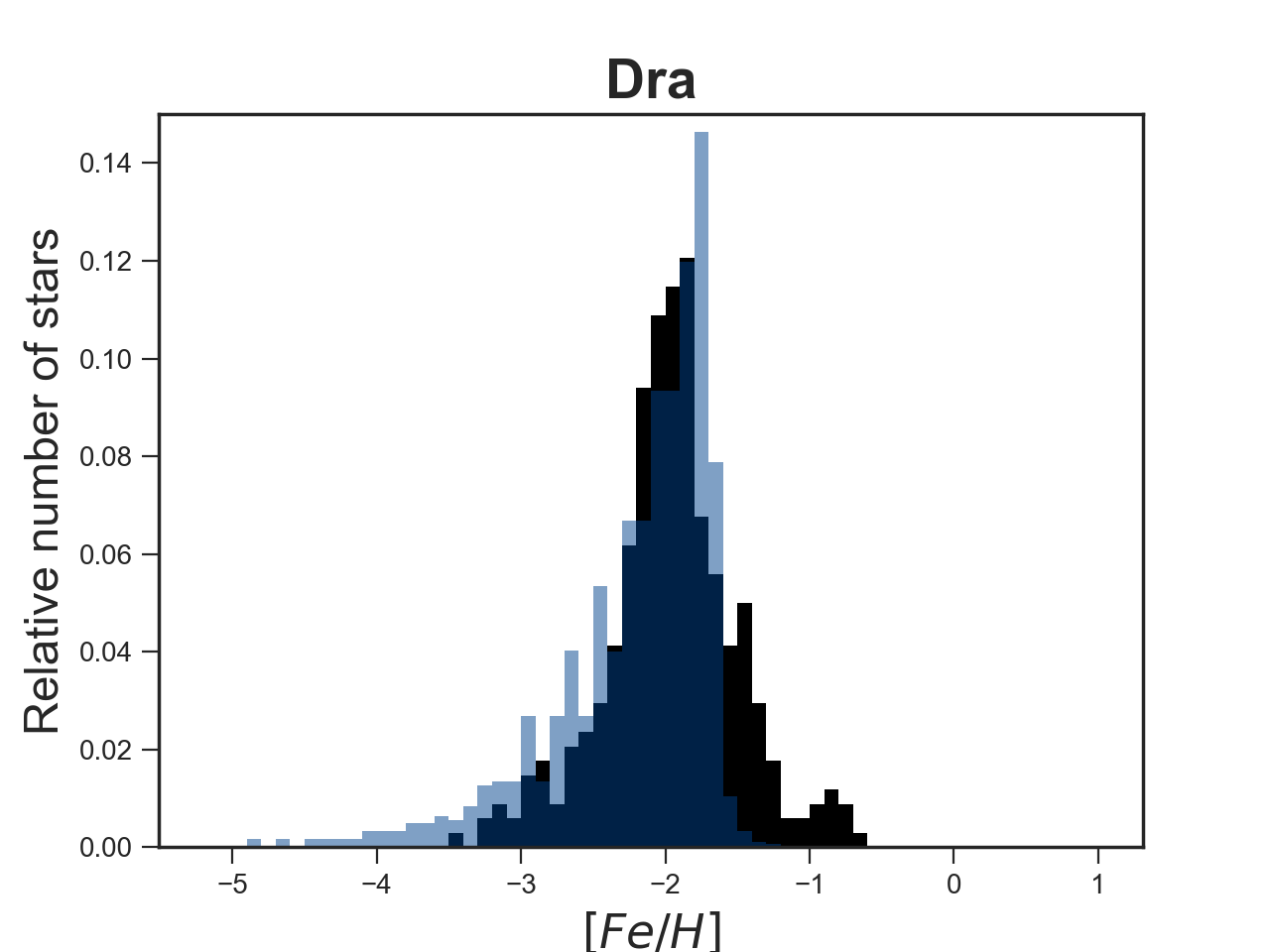}
  \label{fig:3}
\end{subfigure}
\medskip
\begin{subfigure}{0.30\textwidth}
  \includegraphics[width=\linewidth]{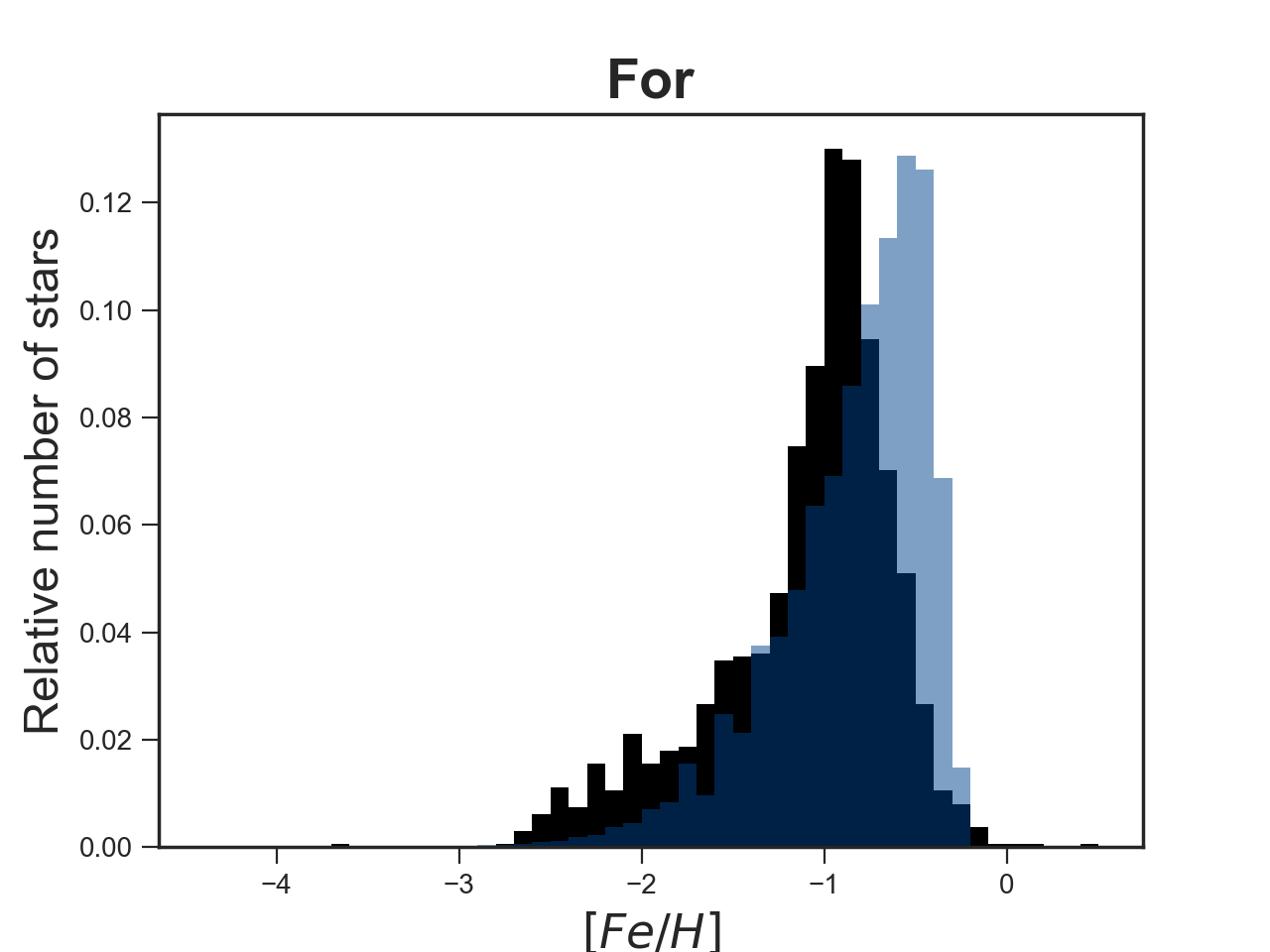}
  \label{fig:4}
\end{subfigure}\hfil
\begin{subfigure}{0.30\textwidth}
  \includegraphics[width=\linewidth]{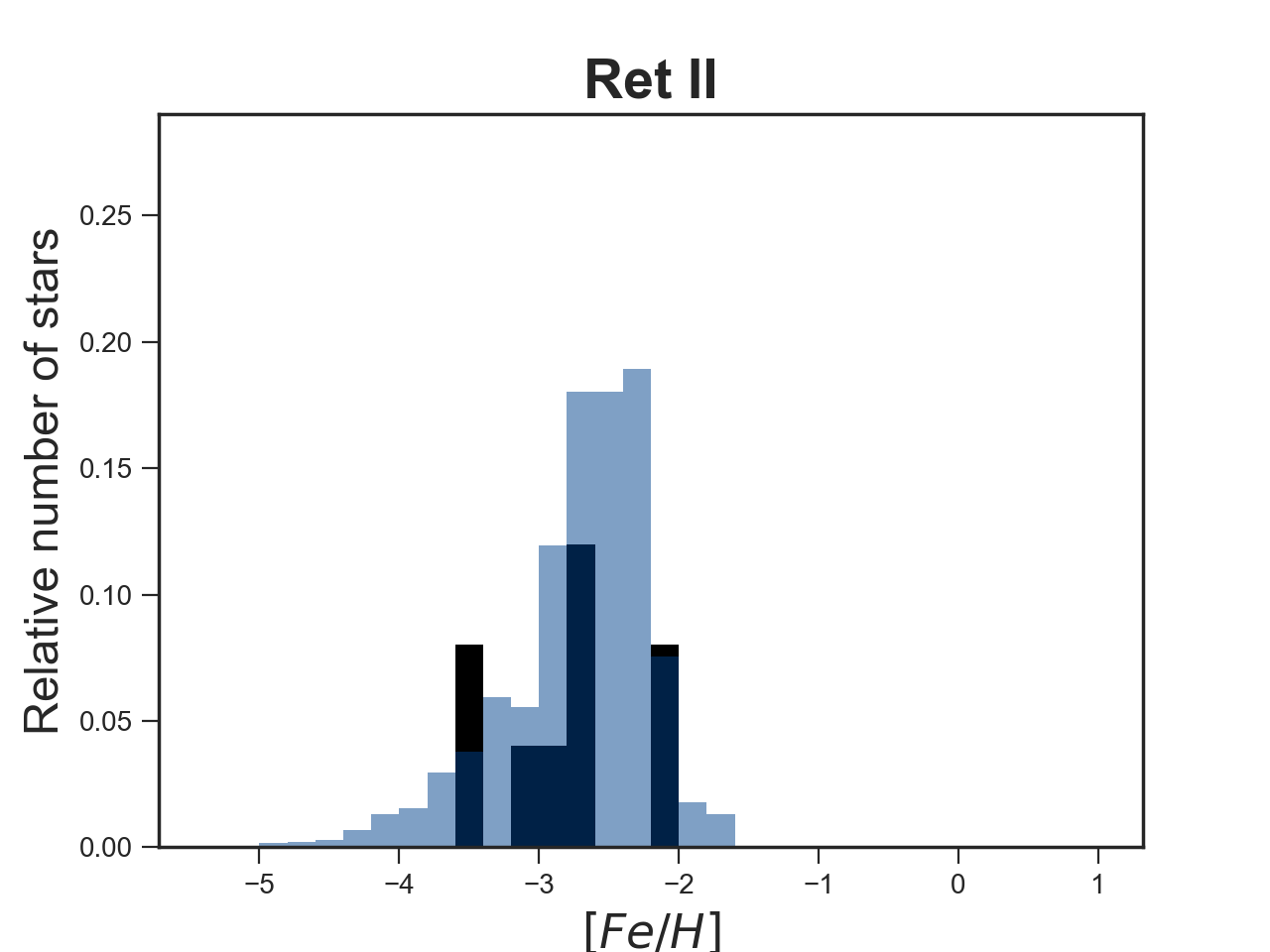}
  \label{fig:5}
\end{subfigure}\hfil 
\begin{subfigure}{0.30\textwidth}
  \includegraphics[width=\linewidth]{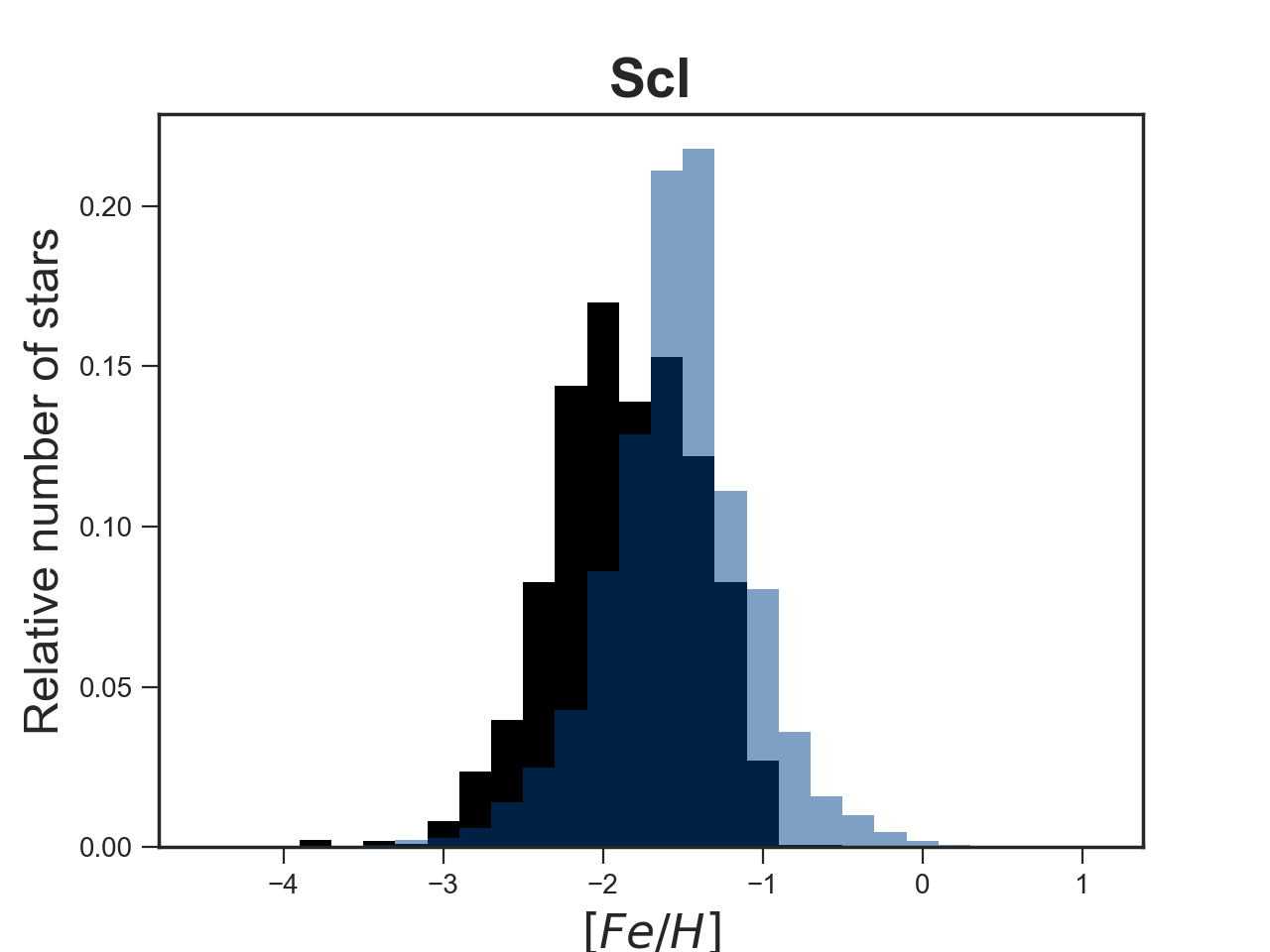}
  \label{fig:6}
\end{subfigure}
\label{fig:images}
\medskip
\begin{subfigure}{0.30\textwidth}
  \includegraphics[width=\linewidth]{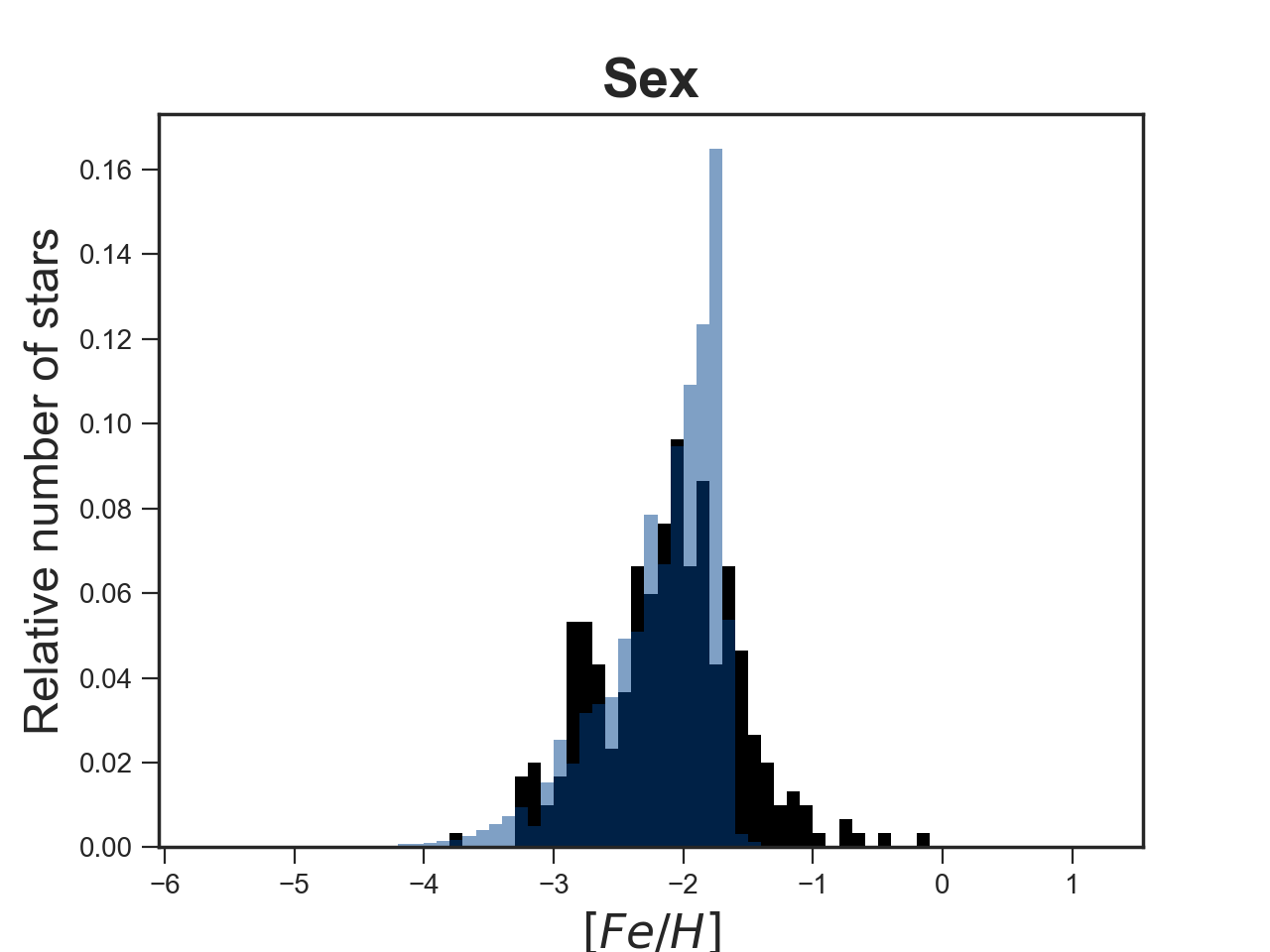}
  \label{fig:7}
\end{subfigure}\hfil 
\begin{subfigure}{0.30\textwidth}
  \includegraphics[width=\linewidth]{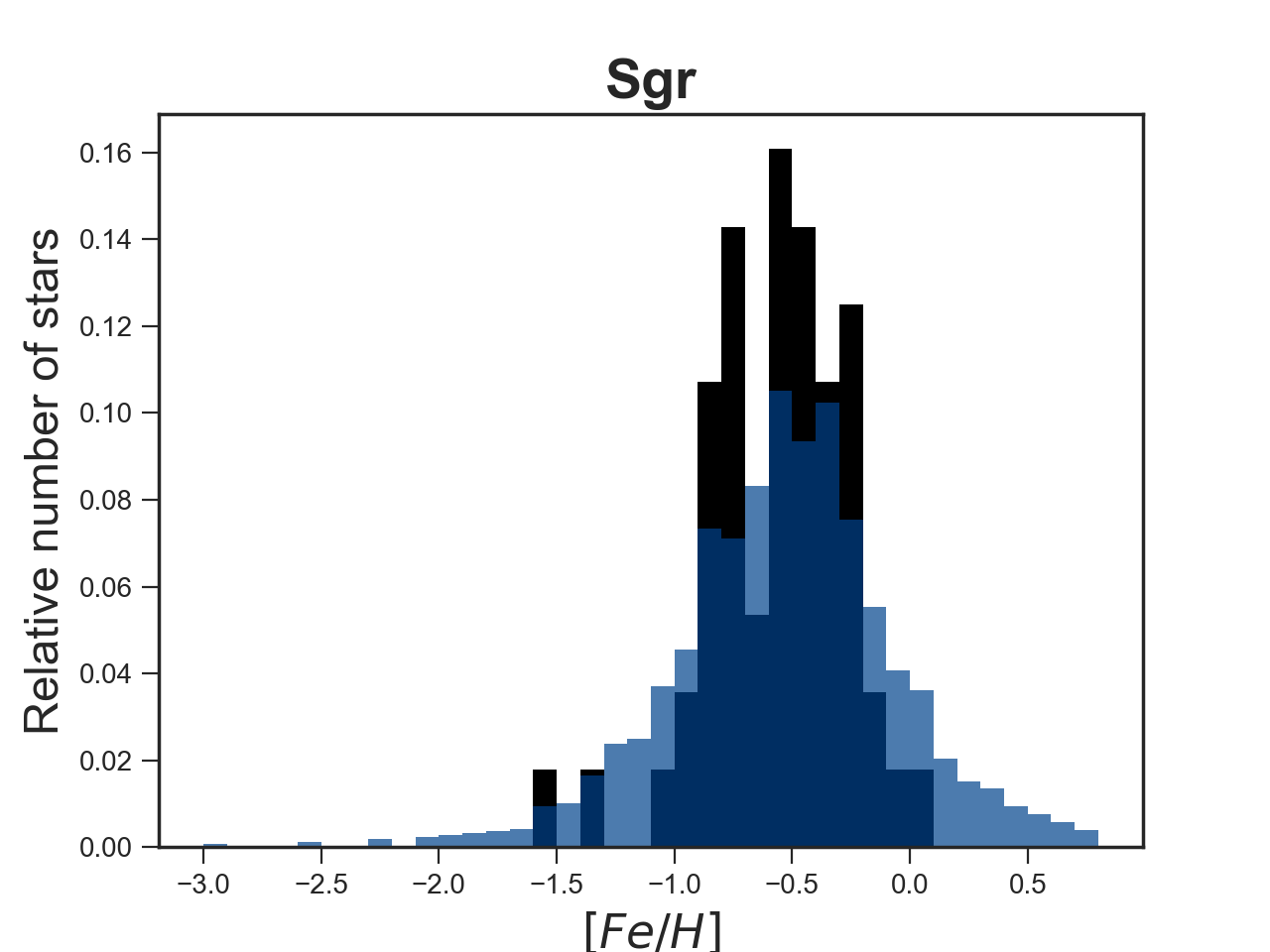}
  \label{fig:8}
\end{subfigure}\hfil 
\begin{subfigure}{0.30\textwidth}
  \includegraphics[width=\linewidth]{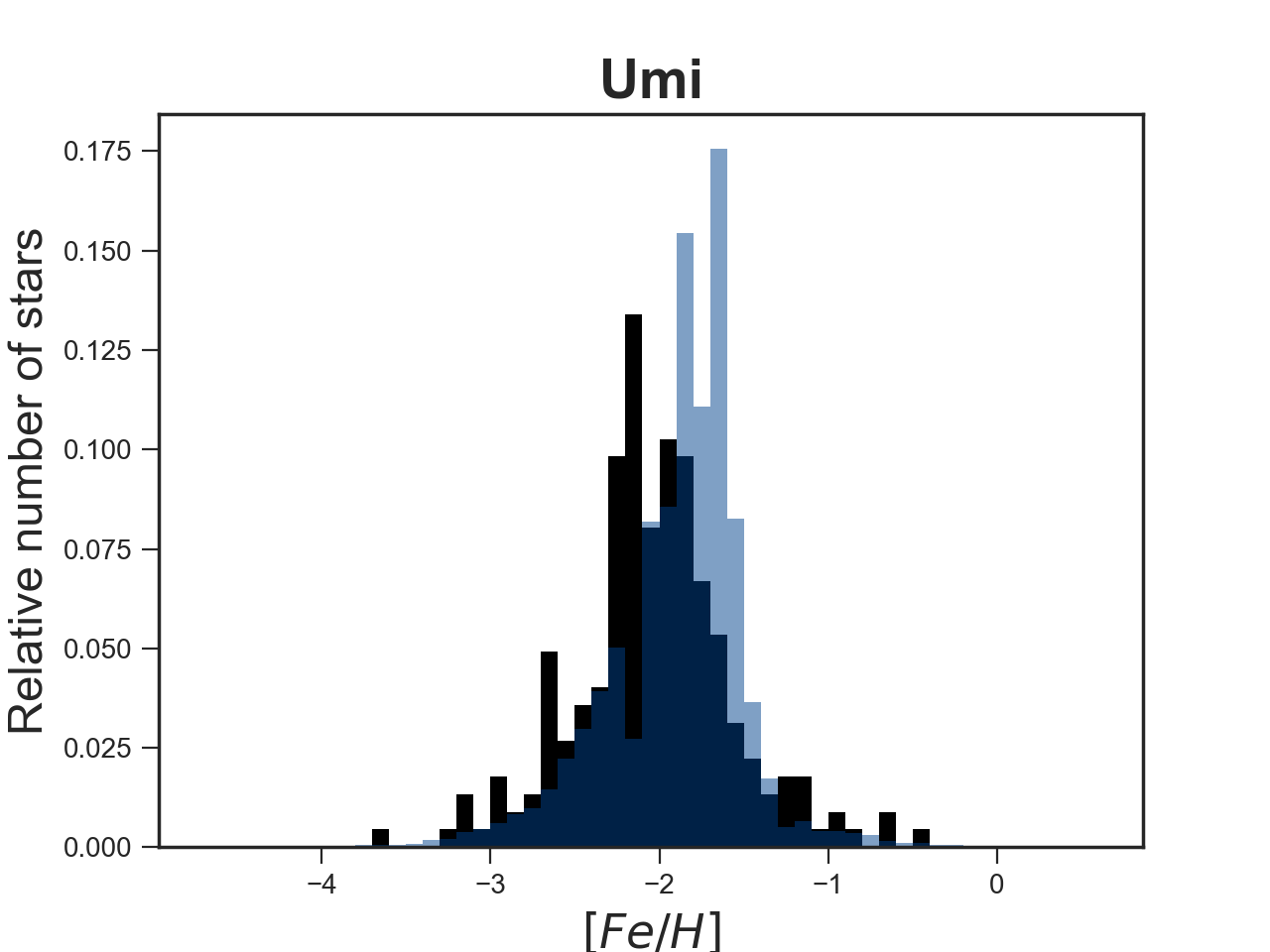}
  \label{fig:9}
\end{subfigure}
\caption{Predicted (light blue) against observed (black) MDF for  Bootes I, Carina, Draco, Fornax, Reticulum II, Sculptor, Sextan, Sagittarius and Ursa Minor I. Observational data adopted for each galaxy are from SAGA database.}
\label{fig:MDF}
\end{figure*}

%\begin{figure*}
%    \centering 
%\begin{subfigure}{0.30\textwidth}
%  \includegraphics[width=\linewidth]{Images/Lithium2/Li2_Boo.png}
%  \label{fig:1}
%\end{subfigure}\hfil 
%\begin{subfigure}{0.30\textwidth}
%  \includegraphics[width=\linewidth]{Images/Lithium2/Li2_Car.png}
%  \label{fig:2}
%\end{subfigure}\hfil 
%\begin{subfigure}{0.30\textwidth}
%  \includegraphics[width=\linewidth]{Images/Lithium2/Li2_Dra.png}
%  \label{fig:3}
%\end{subfigure}
%\medskip
%\begin{subfigure}{0.30\textwidth}
%  \includegraphics[width=\linewidth]{Images/Lithium2/Li2_For.png}
%  \label{fig:4}
%\end{subfigure}\hfil
%\begin{subfigure}{0.30\textwidth}
%  \includegraphics[width=\linewidth]{Images/Lithium2/Li2_Ret.png}
%  \label{fig:5}
%\end{subfigure}\hfil 
%\begin{subfigure}{0.30\textwidth}
%  \includegraphics[width=\linewidth]{Images/Lithium2/Li2_Scl.png}
%  \label{fig:6}
%\end{subfigure}
%\label{fig:images}
%\medskip
%\begin{subfigure}{0.30\textwidth}
%  \includegraphics[width=\linewidth]{Images/Lithium2/Li2_Sex.png}
%  \label{fig:7}
%\end{subfigure}\hfil 
%\begin{subfigure}{0.30\textwidth}
%  \includegraphics[width=\linewidth]{Images/Lithium2/Li2_Sgr.png}
%  \label{fig:8}
%\end{subfigure}\hfil 
%\begin{subfigure}{0.30\textwidth}
%  \includegraphics[width=\linewidth]{Images/Lithium2/Li2_Umi.png}
%  \label{fig:9}
%\end{subfigure}
%\caption{Put the caption here}
%\label{fig:Lithium_2}
%\end{figure*}

\section{Model results}

In Figures \ref{fig:Lithium_Both 1} and \ref{fig: Lithium_Both 2}, we show the predicted evolution of A(Li)=12 +log (Li/H) as a function of [Fe/H] for all the galaxies considered in the present study and compared with data of similar objects. We remind that our aim is fitting only the upper envelope of the data, because Li is easily destroyed at temperatures  $ >\sim 2.8 \cdot 10^{6}$ K. in stars of all masses.
We predict the Li abundance in the ISM of each galaxy and then we compare it with the upper envelope of the stellar data. Most of the stars in these galaxies are confined at metallicities lower than [Fe/H] = -1.0 dex, therefore the values of A(Li) for metallicities larger than that refer mostly to the ISM (thin lines in the Figures). In other words, the Li abundances for [Fe/H]$>-1.0$ dex are mostly predictions, and it appears that dwarf galaxies have the majority of their stars on the Spite plateau (see also Figure \ref{fig:MDF}). If in the future we will be able to measure Li in more metal-rich stars in dwarf galaxies, we should expect to find them in Sagittarius, Carina, Sculptor, Ursa Minor and Fornax, but not in Bootes I and Reticulum, the two UfDs, which, on the other hand, should show only the Spite plateau since they are very metal-poor objects and less evolved than dSphs. In any case, the ISM of all the dwarfs should show the same Li evolution as we observe in the Milky Way, namely the steep rise of Li abundance which is due, in our models, to the late contribution to Li from novae. The AGB stars can also contribute to Li production at late times, but they likely produce negligible amounts of this element (\citealp{Karakas2010, Ventura2000, Ventura2020}).
Another possible late source of Li are the RG stars, recently reappreciated on the basis of GALAH data (\citealp{Kumar2020}), although extreme assumptions on Li production are necessary, as discussed in \cite{Romano2001}. Moreover, theoretical models of low mass stars including rotation and thermohaline mixing have shown that these stars efficiently destroy Li during their lifetime (\citealp{Charbonnel2020}).
From Figures 2 and 3, one can see two curves for each galaxy, one sitting perfectly on the Spite plateau, since we have assumed the Li value of the plateau as the primordial Li abundance, while the other curve is obtained by adopting the SBBN Li primordial abundance. It is worth noting that, as for the Milky Way (see \cite{Izzo2015}), the two theoretical curves merge at high metallicity, ending up with the same maximum value. This is because Li destruction/production by successive stellar generations overwhelms the primordial contribution by infalling gas, thus the system loses memory of its initial Li content.

The observed and theoretical stellar metallicity distribution for the studied galaxies is clearly seen in Figure \ref{fig:MDF}, where we report the predicted and observed MDF for each galaxy. The fit to the observational data in some cases is not perfect and it is mostly due to the lack of observational points, but in other cases such as Carina, Draco and Sculptor the fit is very good.
%Finally, Figure \ref{fig:Lithium_2} is the same as Figure \ref{fig:Lithium_1} but we %reported only the predictions up to the maximum [Fe/H] at which we find stars in these %galaxies. 

\section{Discussion and Conclusions}
We have computed the evolution of $^{7}$Li abundance in several dSphs and UfDs of the Local Group and compared the results with data referring to similar objects and halo stars (\citealp{Aguado2021}). Both models and data show a clear Spite plateau for [Fe/H]$<$ -1.0 dex.
This universality of the Spite plateau is somewhat surprising since we would have expected that the Spite plateau could depend on the environment where stars do form, and therefore to be different in different galaxies which suffered diverse histories of star formation, as it is the case for the sample considered here. In fact, we considered dwarf galaxies, which have suffered  a lower SFR than our Galaxy and also different infalls and outflows of gas.\\
Various explanations have been proposed to explain the Spite plateau in the Milky Way against the higher primordial Li value suggested by BBN. First of all, Li-depletion in stars.
Originally, before WMAP, the common interpretation of the Spite plateau in the Milky Way was that its value corresponds to the primordial Li abundance, and this was justified by the fact that metal-poor stars have thin convective layers and can retain the initial Li abundance of the gas out of which they formed (see \citealp{Deliyannis1990}). In the following years, the difference between the Li value in the Spite plateau and the primordial value from WMAP has been attributed to Li depletion in pre-Main Sequence stars (e.g. \citealp{fu2015}). There is a general consensus now that the Li depletion occurs by means of extra-mixing below the convective zone and that this can be due to rotation, diffusion and turbulent mixing (e.g.\citealp{Richard2005, Richard2012}). Internal gravity waves have also been suggested (\citealp{Charbonnel2005}). Another suggested possibility is that Pop III pregalactic stars might have reprocessed the gas which later formed the first Pop II stars (\citealp{Piau2006, Piau2008}). However, none of these suggested solutions gives a clear answer on why Li should have been depleted by the same amount in all the halo stars as well as in stars of external galaxies and also of Gaia Sausage/Enceladus (\citealp{Molaro2020}). This object has been recognized to be the debris of a merger event occurred in the Milky Way 10 Gyr ago (\citealt{Belokurov2018, Myeong2018, Helmi2018}).
Our conclusions are:
\begin{itemize}
    \item Models of galactic chemical evolution of dSphs and UfDs, can very well reproduce the observed Li abundances, if we assume that the primordial Li abundance coincides with the Spite plateau. This is because we assume that Li is mainly produced in novae, whose contribution appears for ISM metallicities [Fe/H] $>$ -1.0 dex, and very few stars are found at those metallicities in these galaxies. Therefore, we just predict the expected Li abundance for stars with [Fe/H]$>$-1.0 dex and undepleted Li (namely inhabiting the upper envelope of the A(Li) vs. [Fe/H] space).
    
    \item However, since cosmology suggests a different value for primordial Li abundance, which is 2-3 times higher than the Spite plateau, we have also computed models starting from this primordial abundance. The model predictions lie indeed above the Spite plateau for [Fe/H] $<$-1.0 dex but then they merge with the predictions, obtained by adopting the primordial Li of the Spite plateau, for larger metallicities.
    
    \item Very likely the halo of the Milky Way has been partly formed by accretion of stars from dwarf satellites, although we do not know yet how many stars formed in-situ and ex-situ.
    The mixture of in-situ and ex-situ halo stars are all apparently lying on the Spite plateau, so Li is not a good discriminant for the origin of halo stars. The easiest explanation for that, would be that the primordial Li abundance is indeed that of the Spite plateau, supporting a lower primordial Li production driven by non-standard nucleosynthesis processes.  Otherwise a universal, unlikely, Li depletion mechanism, independent of chemical composition and effective temperature should be acting in low metallicity stars. 
    %However, no clear detection of such a warm plateau is evident in their data. 
    In conclusion, the universality of the Spite plateau is not yet understood and more data are required especially at very low metallicities, before drawing firm conclusions.
\end{itemize}

\section*{Acknowledgements}
We thanks an anonymous referee foe carefully reading the manuscript and giving useful suggestions.

\section*{Data Availability}
The data underlying this article will be shared upon request.

%%%%%%%%%%%%%%%%%%%%%%%%%%%%%%%%%%%%%%%%%%%%%%%%%%



%\bibliographystyle{mnras}

\begin{thebibliography}{}
\makeatletter
\relax
\def\mn@urlcharsother{\let\do\@makeother \do\$\do\&\do\#\do\^\do\_\do\%\do\~}
\def\mn@doi{\begingroup\mn@urlcharsother \@ifnextchar [ {\mn@doi@}
  {\mn@doi@[]}}
\def\mn@doi@[#1]#2{\def\@tempa{#1}\ifx\@tempa\@empty \href
  {http://dx.doi.org/#2} {doi:#2}\else \href {http://dx.doi.org/#2} {#1}\fi
  \endgroup}
\def\mn@eprint#1#2{\mn@eprint@#1:#2::\@nil}
\def\mn@eprint@arXiv#1{\href {http://arxiv.org/abs/#1} {{\tt arXiv:#1}}}
\def\mn@eprint@dblp#1{\href {http://dblp.uni-trier.de/rec/bibtex/#1.xml}
  {dblp:#1}}
\def\mn@eprint@#1:#2:#3:#4\@nil{\def\@tempa {#1}\def\@tempb {#2}\def\@tempc
  {#3}\ifx \@tempc \@empty \let \@tempc \@tempb \let \@tempb \@tempa \fi \ifx
  \@tempb \@empty \def\@tempb {arXiv}\fi \@ifundefined
  {mn@eprint@\@tempb}{\@tempb:\@tempc}{\expandafter \expandafter \csname
  mn@eprint@\@tempb\endcsname \expandafter{\@tempc}}}

\bibitem[\protect\citeauthoryear{{Aguado}, {Allende Prieto}, {Gonz{\'a}lez
  Hern{\'a}ndez}  \& {Rebolo}}{{Aguado} et~al.}{2018a}]{Agu18II}
{Aguado} D.~S.,  {Allende Prieto} C.,  {Gonz{\'a}lez Hern{\'a}ndez} J.~I.,
  {Rebolo} R.,  2018a, \mn@doi [\apjl] {10.3847/2041-8213/aaadb8}, \href
  {http://ukads.nottingham.ac.uk/abs/2018ApJ...854L..34A} {854, L34}

\bibitem[\protect\citeauthoryear{{Aguado}, {Gonz{\'a}lez Hern{\'a}ndez},
  {Allende Prieto}  \& {Rebolo}}{{Aguado} et~al.}{2018b}]{Agu18I}
{Aguado} D.~S.,  {Gonz{\'a}lez Hern{\'a}ndez} J.~I.,  {Allende Prieto} C.,
  {Rebolo} R.,  2018b, \mn@doi [ApJL] {10.1086/498131}, 852 L20

\bibitem[\protect\citeauthoryear{{Aguado}, {Gonz{\'a}lez Hern{\'a}ndez},
  {Allende Prieto}  \& {Rebolo}}{{Aguado} et~al.}{2019}]{Aguado2019}
{Aguado} D.~S.,  {Gonz{\'a}lez Hern{\'a}ndez} J.~I.,  {Allende Prieto} C.,
  {Rebolo} R.,  2019, \mn@doi [\apjl] {10.3847/2041-8213/ab1076}, \href
  {https://ui.adsabs.harvard.edu/abs/2019ApJ...874L..21A} {874, L21}

\bibitem[\protect\citeauthoryear{{Aguado} et~al.,}{{Aguado}
  et~al.}{2021}]{Aguado2021}
{Aguado} D.~S.,  et~al., 2021, \mn@doi [\mnras] {10.1093/mnras/staa3250}, \href
  {https://ui.adsabs.harvard.edu/abs/2021MNRAS.500..889A} {500, 889}

\bibitem[\protect\citeauthoryear{{Allende Prieto} et~al.,}{{Allende Prieto}
  et~al.}{2015}]{2015allende}
{Allende Prieto} C.,  et~al., 2015, \mn@doi [\aap]
  {10.1051/0004-6361/201525904}, \href
  {https://ui.adsabs.harvard.edu/abs/2015A&A...579A..98A} {579, A98}

\bibitem[\protect\citeauthoryear{{Bath} \& {Shaviv}}{{Bath} \&
  {Shaviv}}{1978}]{Bath1978}
{Bath} G.~T.,  {Shaviv} G.,  1978, \mn@doi [\mnras] {10.1093/mnras/183.3.515},
  \href {https://ui.adsabs.harvard.edu/abs/1978MNRAS.183..515B} {183, 515}

\bibitem[\protect\citeauthoryear{{Belokurov}, {Erkal}, {Evans}, {Koposov}  \&
  {Deason}}{{Belokurov} et~al.}{2018}]{Belokurov2018}
{Belokurov} V.,  {Erkal} D.,  {Evans} N.~W.,  {Koposov} S.~E.,   {Deason}
  A.~J.,  2018, \mn@doi [\mnras] {10.1093/mnras/sty982}, \href
  {https://ui.adsabs.harvard.edu/abs/2018MNRAS.478..611B} {478, 611}

\bibitem[\protect\citeauthoryear{{Bonifacio} \& {Molaro}}{{Bonifacio} \&
  {Molaro}}{1997}]{Bonifacio1997}
{Bonifacio} P.,  {Molaro} P.,  1997, \mn@doi [\mnras]
  {10.1093/mnras/285.4.847}, \href
  {https://ui.adsabs.harvard.edu/abs/1997MNRAS.285..847B} {285, 847}

\bibitem[\protect\citeauthoryear{{Bonifacio} et~al.,}{{Bonifacio}
  et~al.}{2007}]{Bonifacio2007}
{Bonifacio} P.,  et~al., 2007, \mn@doi [\aap] {10.1051/0004-6361:20064834},
  \href {https://ui.adsabs.harvard.edu/abs/2007A&A...462..851B} {462, 851}

\bibitem[\protect\citeauthoryear{{Bonifacio} et~al.,}{{Bonifacio}
  et~al.}{2015a}]{Bonifacio2015}
{Bonifacio} P.,  et~al., 2015a, \mn@doi [\aap] {10.1051/0004-6361/201425266},
  \href {https://ui.adsabs.harvard.edu/abs/2015A&A...579A..28B} {579, A28}

\bibitem[\protect\citeauthoryear{{Bonifacio} et~al.,}{{Bonifacio}
  et~al.}{2015b}]{boni15}
{Bonifacio} P.,  et~al., 2015b, \mn@doi [\aap] {10.1051/0004-6361/201425266},
  \href {http://adsabs.harvard.edu/abs/2015A%26A...579A..28B} {579, A28}

\bibitem[\protect\citeauthoryear{{Brown} et~al.,}{{Brown}
  et~al.}{2014}]{Brown2014}
{Brown} T.~M.,  et~al., 2014, \mn@doi [\apj] {10.1088/0004-637X/796/2/91},
  \href {https://ui.adsabs.harvard.edu/abs/2014ApJ...796...91B} {796, 91}

\bibitem[\protect\citeauthoryear{{Caffau} et~al.,}{{Caffau}
  et~al.}{2012}]{2012caffau}
{Caffau} E.,  et~al., 2012, \mn@doi [\aap] {10.1051/0004-6361/201118744}, \href
  {https://ui.adsabs.harvard.edu/abs/2012A&A...542A..51C} {542, A51}

\bibitem[\protect\citeauthoryear{{Cameron} \& {Fowler}}{{Cameron} \&
  {Fowler}}{1971}]{Cameron1971}
{Cameron} A.~G.~W.,  {Fowler} W.~A.,  1971, \mn@doi [\apj] {10.1086/150821},
  \href {https://ui.adsabs.harvard.edu/abs/1971ApJ...164..111C} {164, 111}

\bibitem[\protect\citeauthoryear{{Charbonnel} \& {Talon}}{{Charbonnel} \&
  {Talon}}{2005}]{Charbonnel2005}
{Charbonnel} C.,  {Talon} S.,  2005, in {Alecian} G.,  {Richard} O.,
  {Vauclair} S.,  eds,  EAS Publications Series Vol. 17, EAS Publications
  Series. pp 167--176, \mn@doi{10.1051/eas:2005111}

\bibitem[\protect\citeauthoryear{{Charbonnel} et~al.,}{{Charbonnel}
  et~al.}{2020}]{Charbonnel2020}
{Charbonnel} C.,  et~al., 2020, \mn@doi [\aap] {10.1051/0004-6361/201936360},
  \href {https://ui.adsabs.harvard.edu/abs/2020A&A...633A..34C} {633, A34}

\bibitem[\protect\citeauthoryear{{Christlieb}, {Gustafsson}, {Korn}, {Barklem},
  {Beers}, {Bessell}, {Karlsson}  \& {Mizuno-Wiedner}}{{Christlieb}
  et~al.}{2004}]{chris04}
{Christlieb} N.,  {Gustafsson} B.,  {Korn} A.~J.,  {Barklem} P.~S.,  {Beers}
  T.~C.,  {Bessell} M.~S.,  {Karlsson} T.,   {Mizuno-Wiedner} M.,  2004,
  \mn@doi [\apj] {10.1086/381237}, \href
  {http://adsabs.harvard.edu/abs/2004ApJ...603..708C} {603, 708}

\bibitem[\protect\citeauthoryear{{Coc}, {Uzan}  \& {Vangioni}}{{Coc}
  et~al.}{2014}]{Coc2014}
{Coc} A.,  {Uzan} J.-P.,   {Vangioni} E.,  2014, \mn@doi [\jcap]
  {10.1088/1475-7516/2014/10/050}, \href
  {https://ui.adsabs.harvard.edu/abs/2014JCAP...10..050C} {2014, 050}

\bibitem[\protect\citeauthoryear{{D'Antona} \& {Matteucci}}{{D'Antona} \&
  {Matteucci}}{1991}]{Dantona1991}
{D'Antona} F.,  {Matteucci} F.,  1991, \aap, \href
  {https://ui.adsabs.harvard.edu/abs/1991A&A...248...62D} {248, 62}

\bibitem[\protect\citeauthoryear{{Deliyannis}, {Demarque}  \&
  {Kawaler}}{{Deliyannis} et~al.}{1990}]{Deliyannis1990}
{Deliyannis} C.~P.,  {Demarque} P.,   {Kawaler} S.~D.,  1990, \mn@doi [\apjs]
  {10.1086/191439}, \href
  {https://ui.adsabs.harvard.edu/abs/1990ApJS...73...21D} {73, 21}

\bibitem[\protect\citeauthoryear{{Doherty}, {Gil-Pons}, {Lau}, {Lattanzio}  \&
  {Siess}}{{Doherty} et~al.}{2014a}]{Doherty20142}
{Doherty} C.~L.,  {Gil-Pons} P.,  {Lau} H. H.~B.,  {Lattanzio} J.~C.,   {Siess}
  L.,  2014a, \mn@doi [\mnras] {10.1093/mnras/stt1877}, \href
  {https://ui.adsabs.harvard.edu/abs/2014MNRAS.437..195D} {437, 195}

\bibitem[\protect\citeauthoryear{{Doherty}, {Gil-Pons}, {Lau}, {Lattanzio},
  {Siess}  \& {Campbell}}{{Doherty} et~al.}{2014b}]{Doherty2014}
{Doherty} C.~L.,  {Gil-Pons} P.,  {Lau} H. H.~B.,  {Lattanzio} J.~C.,  {Siess}
  L.,   {Campbell} S.~W.,  2014b, \mn@doi [\mnras] {10.1093/mnras/stu571},
  \href {https://ui.adsabs.harvard.edu/abs/2014MNRAS.441..582D} {441, 582}

\bibitem[\protect\citeauthoryear{{Dolphin}}{{Dolphin}}{2002}]{Dolphin2002}
{Dolphin} A.~E.,  2002, \mn@doi [\mnras] {10.1046/j.1365-8711.2002.05271.x},
  \href {https://ui.adsabs.harvard.edu/abs/2002MNRAS.332...91D} {332, 91}

\bibitem[\protect\citeauthoryear{{Frebel} et~al.,}{{Frebel}
  et~al.}{2005}]{fre05}
{Frebel} A.,  et~al., 2005, \mn@doi [\nat] {10.1038/nature03455}, \href
  {http://adsabs.harvard.edu/abs/2005Natur.434..871F} {434, 871}

\bibitem[\protect\citeauthoryear{{Fu}, {Bressan}, {Molaro}  \& {Marigo}}{{Fu}
  et~al.}{2015}]{fu2015}
{Fu} X.,  {Bressan} A.,  {Molaro} P.,   {Marigo} P.,  2015, \mn@doi [\mnras]
  {10.1093/mnras/stv1384}, \href
  {https://ui.adsabs.harvard.edu/abs/2015MNRAS.452.3256F} {452, 3256}

\bibitem[\protect\citeauthoryear{{Gao} et~al.,}{{Gao} et~al.}{2020}]{Gao2020}
{Gao} X.,  et~al., 2020, \mn@doi [\mnras] {10.1093/mnrasl/slaa109}, \href
  {https://ui.adsabs.harvard.edu/abs/2020MNRAS.497L..30G} {497, L30}

\bibitem[\protect\citeauthoryear{{Grisoni}, {Matteucci}, {Romano}  \&
  {Fu}}{{Grisoni} et~al.}{2019}]{Grisoni2019}
{Grisoni} V.,  {Matteucci} F.,  {Romano} D.,   {Fu} X.,  2019, \mn@doi [\mnras]
  {10.1093/mnras/stz2428}, \href
  {https://ui.adsabs.harvard.edu/abs/2019MNRAS.489.3539G} {489, 3539}

\bibitem[\protect\citeauthoryear{{Hansen} et~al.,}{{Hansen}
  et~al.}{2014}]{2014hansen}
{Hansen} T.,  et~al., 2014, \mn@doi [\apj] {10.1088/0004-637X/787/2/162}, \href
  {https://ui.adsabs.harvard.edu/abs/2014ApJ...787..162H} {787, 162}

\bibitem[\protect\citeauthoryear{{Helmi}, {Babusiaux}, {Koppelman}, {Massari},
  {Veljanoski}  \& {Brown}}{{Helmi} et~al.}{2018}]{Helmi2018}
{Helmi} A.,  {Babusiaux} C.,  {Koppelman} H.~H.,  {Massari} D.,  {Veljanoski}
  J.,   {Brown} A. G.~A.,  2018, \mn@doi [\nat] {10.1038/s41586-018-0625-x},
  \href {https://ui.adsabs.harvard.edu/abs/2018Natur.563...85H} {563, 85}

\bibitem[\protect\citeauthoryear{{Hernandez}, {Gilmore}  \&
  {Valls-Gabaud}}{{Hernandez} et~al.}{2000}]{Hernandez2000}
{Hernandez} X.,  {Gilmore} G.,   {Valls-Gabaud} D.,  2000, \mn@doi [\mnras]
  {10.1046/j.1365-8711.2000.03809.x}, \href
  {https://ui.adsabs.harvard.edu/abs/2000MNRAS.317..831H} {317, 831}

\bibitem[\protect\citeauthoryear{{Hill} et~al.,}{{Hill}
  et~al.}{2019}]{Hill2019}
{Hill} V.,  et~al., 2019, \mn@doi [\aap] {10.1051/0004-6361/201833950}, \href
  {https://ui.adsabs.harvard.edu/abs/2019A&A...626A..15H} {626, A15}

\bibitem[\protect\citeauthoryear{{Hisano}, {Kawasaki}, {Kohri}, {Moroi}  \&
  {Nakayama}}{{Hisano} et~al.}{2009}]{Hisano2009}
{Hisano} J.,  {Kawasaki} M.,  {Kohri} K.,  {Moroi} T.,   {Nakayama} K.,  2009,
  \mn@doi [\prd] {10.1103/PhysRevD.79.083522}, \href
  {https://ui.adsabs.harvard.edu/abs/2009PhRvD..79h3522H} {79, 083522}

\bibitem[\protect\citeauthoryear{{Izzo} et~al.,}{{Izzo}
  et~al.}{2015}]{Izzo2015}
{Izzo} L.,  et~al., 2015, \mn@doi [\apjl] {10.1088/2041-8205/808/1/L14}, \href
  {https://ui.adsabs.harvard.edu/abs/2015ApJ...808L..14I} {808, L14}

\bibitem[\protect\citeauthoryear{{Jedamzik}, {Choi}, {Roszkowski}  \& {Ruiz de
  Austri}}{{Jedamzik} et~al.}{2006}]{Jedamzik2006}
{Jedamzik} K.,  {Choi} K.-Y.,  {Roszkowski} L.,   {Ruiz de Austri} R.,  2006,
  \mn@doi [\jcap] {10.1088/1475-7516/2006/07/007}, \href
  {https://ui.adsabs.harvard.edu/abs/2006JCAP...07..007J} {2006, 007}

\bibitem[\protect\citeauthoryear{{Jos{\'e}} \& {Hernanz}}{{Jos{\'e}} \&
  {Hernanz}}{2007}]{Jose2007}
{Jos{\'e}} J.,  {Hernanz} M.,  2007, \mn@doi [Journal of Physics G Nuclear
  Physics] {10.1088/0954-3899/34/12/R01}, \href
  {https://ui.adsabs.harvard.edu/abs/2007JPhG...34..431J} {34, R431}

\bibitem[\protect\citeauthoryear{{Karakas}}{{Karakas}}{2010}]{Karakas2010}
{Karakas} A.~I.,  2010, \mn@doi [\mnras] {10.1111/j.1365-2966.2009.16198.x},
  \href {https://ui.adsabs.harvard.edu/abs/2010MNRAS.403.1413K} {403, 1413}

\bibitem[\protect\citeauthoryear{{Keller} et~al.,}{{Keller}
  et~al.}{2014}]{kel14}
{Keller} S.~C.,  et~al., 2014, \mn@doi [\nat] {10.1038/nature12990}, \href
  {http://adsabs.harvard.edu/abs/2014Natur.506..463K} {506, 463}

\bibitem[\protect\citeauthoryear{{Kennicutt}}{{Kennicutt}}{1998}]{kennicutt}
{Kennicutt} Robert~C. J.,  1998, \mn@doi [\apj] {10.1086/305588}, \href
  {https://ui.adsabs.harvard.edu/abs/1998ApJ...498..541K} {498, 541}

\bibitem[\protect\citeauthoryear{{Kumar}, {Reddy}, {Campbell}, {Maben}, {Zhao}
  \& {Ting}}{{Kumar} et~al.}{2020}]{Kumar2020}
{Kumar} Y.~B.,  {Reddy} B.~E.,  {Campbell} S.~W.,  {Maben} S.,  {Zhao} G.,
  {Ting} Y.-S.,  2020, \mn@doi [Nature Astronomy] {10.1038/s41550-020-1139-7},
  \href {https://ui.adsabs.harvard.edu/abs/2020NatAs...4.1059K} {4, 1059}

\bibitem[\protect\citeauthoryear{{Lanfranchi} \& {Matteucci}}{{Lanfranchi} \&
  {Matteucci}}{2003}]{Lanfranchi2003}
{Lanfranchi} G.~A.,  {Matteucci} F.,  2003, \mn@doi [\mnras]
  {10.1046/j.1365-8711.2003.06919.x}, \href
  {https://ui.adsabs.harvard.edu/abs/2003MNRAS.345...71L} {345, 71}

\bibitem[\protect\citeauthoryear{{Lanfranchi} \& {Matteucci}}{{Lanfranchi} \&
  {Matteucci}}{2004}]{Lanfranchi&Matteucci04}
{Lanfranchi} G.~A.,  {Matteucci} F.,  2004, \mn@doi [\mnras]
  {10.1111/j.1365-2966.2004.07877.x}, \href
  {https://ui.adsabs.harvard.edu/abs/2004MNRAS.351.1338L} {351, 1338}

\bibitem[\protect\citeauthoryear{{Lemoine}, {Vangioni-Flam}  \&
  {Cass{\'e}}}{{Lemoine} et~al.}{1998}]{Lemoine1998}
{Lemoine} M.,  {Vangioni-Flam} E.,   {Cass{\'e}} M.,  1998, \mn@doi [\apj]
  {10.1086/305650}, \href
  {https://ui.adsabs.harvard.edu/abs/1998ApJ...499..735L} {499, 735}

\bibitem[\protect\citeauthoryear{{Mathews}, {Alcock}  \& {Fuller}}{{Mathews}
  et~al.}{1990}]{Mathews1990}
{Mathews} G.~J.,  {Alcock} C.~R.,   {Fuller} G.~M.,  1990, \mn@doi [\apj]
  {10.1086/168329}, \href
  {https://ui.adsabs.harvard.edu/abs/1990ApJ...349..449M} {349, 449}

\bibitem[\protect\citeauthoryear{{Matteucci}}{{Matteucci}}{2012}]{Matteucci2012}
{Matteucci} F.,  2012, {Chemical Evolution of Galaxies},
  \mn@doi{10.1007/978-3-642-22491-1.
}

\bibitem[\protect\citeauthoryear{{Matteucci}, {D'Antona}  \&
  {Timmes}}{{Matteucci} et~al.}{1995}]{Matteucci1995}
{Matteucci} F.,  {D'Antona} F.,   {Timmes} F.~X.,  1995, \aap, \href
  {https://ui.adsabs.harvard.edu/abs/1995A&A...303..460M} {303, 460}

\bibitem[\protect\citeauthoryear{{Mel{\'e}ndez} et~al.,}{{Mel{\'e}ndez}
  et~al.}{2010}]{Melendez2010}
{Mel{\'e}ndez} J.,  et~al., 2010, \mn@doi [\apss] {10.1007/s10509-009-0187-3},
  \href {https://ui.adsabs.harvard.edu/abs/2010Ap&SS.328..193M} {328, 193}

\bibitem[\protect\citeauthoryear{{Molaro}, {Bressan}, {Barbieri}, {Marigo}  \&
  {Zaggia}}{{Molaro} et~al.}{2012}]{Molaro2012}
{Molaro} P.,  {Bressan} A.,  {Barbieri} M.,  {Marigo} P.,   {Zaggia} S.,  2012,
  Memorie della Societa Astronomica Italiana Supplementi, \href
  {https://ui.adsabs.harvard.edu/abs/2012MSAIS..22..233M} {22, 233}

\bibitem[\protect\citeauthoryear{{Molaro}, {Cescutti}  \& {Fu}}{{Molaro}
  et~al.}{2020}]{Molaro2020}
{Molaro} P.,  {Cescutti} G.,   {Fu} X.,  2020, \mn@doi [\mnras]
  {10.1093/mnras/staa1653}, \href
  {https://ui.adsabs.harvard.edu/abs/2020MNRAS.496.2902M} {496, 2902}

\bibitem[\protect\citeauthoryear{{Monaco}, {Bonifacio}, {Sbordone}, {Villanova}
   \& {Pancino}}{{Monaco} et~al.}{2010}]{Monaco2010}
{Monaco} L.,  {Bonifacio} P.,  {Sbordone} L.,  {Villanova} S.,   {Pancino} E.,
  2010, \mn@doi [\aap] {10.1051/0004-6361/201015162}, \href
  {https://ui.adsabs.harvard.edu/abs/2010A&A...519L...3M} {519, L3}

\bibitem[\protect\citeauthoryear{{Mucciarelli}, {Salaris}, {Bonifacio},
  {Monaco}  \& {Villanova}}{{Mucciarelli} et~al.}{2014}]{Mucciarelli2014}
{Mucciarelli} A.,  {Salaris} M.,  {Bonifacio} P.,  {Monaco} L.,   {Villanova}
  S.,  2014, The Messenger, \href
  {https://ui.adsabs.harvard.edu/abs/2014Msngr.158...45M} {158, 45}

\bibitem[\protect\citeauthoryear{{Myeong}, {Evans}, {Belokurov}, {Sanders}  \&
  {Koposov}}{{Myeong} et~al.}{2018}]{Myeong2018}
{Myeong} G.~C.,  {Evans} N.~W.,  {Belokurov} V.,  {Sanders} J.~L.,   {Koposov}
  S.~E.,  2018, \mn@doi [\apjl] {10.3847/2041-8213/aad7f7}, \href
  {https://ui.adsabs.harvard.edu/abs/2018ApJ...863L..28M} {863, L28}

\bibitem[\protect\citeauthoryear{{Nordlander} et~al.,}{{Nordlander}
  et~al.}{2019}]{nord19}
{Nordlander} T.,  et~al., 2019, \mn@doi [\mnras] {10.1093/mnrasl/slz109}, \href
  {https://ui.adsabs.harvard.edu/abs/2019MNRAS.488L.109N} {488, L109}

\bibitem[\protect\citeauthoryear{{Piau}}{{Piau}}{2008}]{Piau2008}
{Piau} L.,  2008, \mn@doi [\apj] {10.1086/592722}, \href
  {https://ui.adsabs.harvard.edu/abs/2008ApJ...689.1279P} {689, 1279}

\bibitem[\protect\citeauthoryear{{Piau}, {Beers}, {Balsara}, {Sivarani},
  {Truran}  \& {Ferguson}}{{Piau} et~al.}{2006}]{Piau2006}
{Piau} L.,  {Beers} T.~C.,  {Balsara} D.~S.,  {Sivarani} T.,  {Truran} J.~W.,
  {Ferguson} J.~W.,  2006, \mn@doi [\apj] {10.1086/508445}, \href
  {https://ui.adsabs.harvard.edu/abs/2006ApJ...653..300P} {653, 300}

\bibitem[\protect\citeauthoryear{{Prantzos}, {de Laverny}, {Guiglion},
  {Recio-Blanco}  \& {Worley}}{{Prantzos} et~al.}{2017}]{Prantzos2017}
{Prantzos} N.,  {de Laverny} P.,  {Guiglion} G.,  {Recio-Blanco} A.,   {Worley}
  C.~C.,  2017, \mn@doi [\aap] {10.1051/0004-6361/201731188}, \href
  {https://ui.adsabs.harvard.edu/abs/2017A&A...606A.132P} {606, A132}

\bibitem[\protect\citeauthoryear{{Rebolo}, {Molaro}  \& {Beckman}}{{Rebolo}
  et~al.}{1988}]{Rebolo1988}
{Rebolo} R.,  {Molaro} P.,   {Beckman} J.~E.,  1988, \aap, \href
  {https://ui.adsabs.harvard.edu/abs/1988A&A...192..192R} {192, 192}

\bibitem[\protect\citeauthoryear{{Richard}}{{Richard}}{2012}]{Richard2012}
{Richard} O.,  2012, Memorie della Societa Astronomica Italiana Supplementi,
  \href {https://ui.adsabs.harvard.edu/abs/2012MSAIS..22..211R} {22, 211}

\bibitem[\protect\citeauthoryear{{Richard}, {Michaud}  \& {Richer}}{{Richard}
  et~al.}{2005}]{Richard2005}
{Richard} O.,  {Michaud} G.,   {Richer} J.,  2005, \mn@doi [\apj]
  {10.1086/426470}, \href
  {https://ui.adsabs.harvard.edu/abs/2005ApJ...619..538R} {619, 538}

\bibitem[\protect\citeauthoryear{{Romano}, {Matteucci}, {Molaro}  \&
  {Bonifacio}}{{Romano} et~al.}{1999}]{Romano1999}
{Romano} D.,  {Matteucci} F.,  {Molaro} P.,   {Bonifacio} P.,  1999, \aap,
  \href {https://ui.adsabs.harvard.edu/abs/1999A&A...352..117R} {352, 117}

\bibitem[\protect\citeauthoryear{{Romano}, {Matteucci}, {Ventura}  \&
  {D'Antona}}{{Romano} et~al.}{2001}]{Romano2001}
{Romano} D.,  {Matteucci} F.,  {Ventura} P.,   {D'Antona} F.,  2001, \mn@doi
  [\aap] {10.1051/0004-6361:20010751}, \href
  {https://ui.adsabs.harvard.edu/abs/2001A&A...374..646R} {374, 646}

\bibitem[\protect\citeauthoryear{{Romano}, {Karakas}, {Tosi}  \&
  {Matteucci}}{{Romano} et~al.}{2010}]{Romano2010}
{Romano} D.,  {Karakas} A.~I.,  {Tosi} M.,   {Matteucci} F.,  2010, \mn@doi
  [\aap] {10.1051/0004-6361/201014483}, \href
  {https://ui.adsabs.harvard.edu/abs/2010A&A...522A..32R} {522, A32}

\bibitem[\protect\citeauthoryear{{Romano}, {Bellazzini}, {Starkenburg}  \&
  {Leaman}}{{Romano} et~al.}{2015}]{Romano2015}
{Romano} D.,  {Bellazzini} M.,  {Starkenburg} E.,   {Leaman} R.,  2015, \mn@doi
  [\mnras] {10.1093/mnras/stu2427}, \href
  {https://ui.adsabs.harvard.edu/abs/2015MNRAS.446.4220R} {446, 4220}

\bibitem[\protect\citeauthoryear{{Ryan}}{{Ryan}}{2000}]{Ryan2000}
{Ryan} S.~G.,  2000, arXiv e-prints, \href
  {https://ui.adsabs.harvard.edu/abs/2000astro.ph..1230R} {pp
  astro--ph/0001230}

\bibitem[\protect\citeauthoryear{{Salpeter}}{{Salpeter}}{1955}]{salpeter}
{Salpeter} E.~E.,  1955, \mn@doi [\apj] {10.1086/145971}, \href
  {https://ui.adsabs.harvard.edu/abs/1955ApJ...121..161S} {121, 161}

\bibitem[\protect\citeauthoryear{{Salvadori} \& {Ferrara}}{{Salvadori} \&
  {Ferrara}}{2009}]{Salvadori2009}
{Salvadori} S.,  {Ferrara} A.,  2009, \mn@doi [\mnras]
  {10.1111/j.1745-3933.2009.00627.x}, \href
  {https://ui.adsabs.harvard.edu/abs/2009MNRAS.395L...6S} {395, L6}

\bibitem[\protect\citeauthoryear{{Sbordone} et~al.,}{{Sbordone}
  et~al.}{2010}]{Sbordone2010}
{Sbordone} L.,  et~al., 2010, \mn@doi [\aap] {10.1051/0004-6361/200913282},
  \href {https://ui.adsabs.harvard.edu/abs/2010A&A...522A..26S} {522, A26}

\bibitem[\protect\citeauthoryear{{Schmidt}}{{Schmidt}}{1963}]{schmidt}
{Schmidt} M.,  1963, \mn@doi [\apj] {10.1086/147553}, \href
  {https://ui.adsabs.harvard.edu/abs/1963ApJ...137..758S} {137, 758}

\bibitem[\protect\citeauthoryear{{Spergel} et~al.,}{{Spergel}
  et~al.}{2003}]{Spergel2003}
{Spergel} D.~N.,  et~al., 2003, \mn@doi [\apjs] {10.1086/377226}, \href
  {https://ui.adsabs.harvard.edu/abs/2003ApJS..148..175S} {148, 175}

\bibitem[\protect\citeauthoryear{{Spiegel} \& {Zahn}}{{Spiegel} \&
  {Zahn}}{1992}]{Spiegel1992}
{Spiegel} E.~A.,  {Zahn} J.~P.,  1992, \aap, \href
  {https://ui.adsabs.harvard.edu/abs/1992A&A...265..106S} {265, 106}

\bibitem[\protect\citeauthoryear{{Spite} \& {Spite}}{{Spite} \&
  {Spite}}{1982}]{Spite1982}
{Spite} F.,  {Spite} M.,  1982, \aap, \href
  {https://ui.adsabs.harvard.edu/abs/1982A&A...115..357S} {115, 357}

\bibitem[\protect\citeauthoryear{{Spite} \& {Spite}}{{Spite} \&
  {Spite}}{1986}]{Spite1986}
{Spite} F.,  {Spite} M.,  1986, \aap, \href
  {https://ui.adsabs.harvard.edu/abs/1986A&A...163..140S} {163, 140}

\bibitem[\protect\citeauthoryear{{Starkenburg} et~al.,}{{Starkenburg}
  et~al.}{2018}]{2018starkenburg}
{Starkenburg} E.,  et~al., 2018, \mn@doi [\mnras] {10.1093/mnras/sty2276},
  \href {https://ui.adsabs.harvard.edu/abs/2018MNRAS.481.3838S} {481, 3838}

\bibitem[\protect\citeauthoryear{{Starrfield}, {Truran}, {Sparks}  \&
  {Arnould}}{{Starrfield} et~al.}{1978}]{Starrfield1978}
{Starrfield} S.,  {Truran} J.~W.,  {Sparks} W.~M.,   {Arnould} M.,  1978,
  \mn@doi [\apj] {10.1086/156175}, \href
  {https://ui.adsabs.harvard.edu/abs/1978ApJ...222..600S} {222, 600}

\bibitem[\protect\citeauthoryear{{Travaglio}, {Randich}, {Galli}, {Lattanzio},
  {Elliott}, {Forestini}  \& {Ferrini}}{{Travaglio}
  et~al.}{2001}]{Travaglio2001}
{Travaglio} C.,  {Randich} S.,  {Galli} D.,  {Lattanzio} J.,  {Elliott} L.~M.,
  {Forestini} M.,   {Ferrini} F.,  2001, \mn@doi [\apj] {10.1086/322415}, \href
  {https://ui.adsabs.harvard.edu/abs/2001ApJ...559..909T} {559, 909}

\bibitem[\protect\citeauthoryear{{Ventura}, {D'Antona}  \&
  {Mazzitelli}}{{Ventura} et~al.}{2000}]{Ventura2000}
{Ventura} P.,  {D'Antona} F.,   {Mazzitelli} I.,  2000, \aap, \href
  {https://ui.adsabs.harvard.edu/abs/2000A&A...363..605V} {363, 605}

\bibitem[\protect\citeauthoryear{{Ventura}, {Dell'Agli}, {Lugaro}, {Romano},
  {Tailo}  \& {Yag{\"u}e}}{{Ventura} et~al.}{2020}]{Ventura2020}
{Ventura} P.,  {Dell'Agli} F.,  {Lugaro} M.,  {Romano} D.,  {Tailo} M.,
  {Yag{\"u}e} A.,  2020, \mn@doi [\aap] {10.1051/0004-6361/202038289}, \href
  {https://ui.adsabs.harvard.edu/abs/2020A&A...641A.103V} {641, A103}

\bibitem[\protect\citeauthoryear{{Vincenzo}, {Matteucci}, {Vattakunnel}  \&
  {Lanfranchi}}{{Vincenzo} et~al.}{2014}]{Vincenzo2014}
{Vincenzo} F.,  {Matteucci} F.,  {Vattakunnel} S.,   {Lanfranchi} G.~A.,  2014,
  \mn@doi [\mnras] {10.1093/mnras/stu710}, \href
  {https://ui.adsabs.harvard.edu/abs/2014MNRAS.441.2815V} {441, 2815}

\bibitem[\protect\citeauthoryear{{Woosley}, {Hartmann}, {Hoffman}  \&
  {Haxton}}{{Woosley} et~al.}{1990}]{Woosley1990}
{Woosley} S.~E.,  {Hartmann} D.~H.,  {Hoffman} R.~D.,   {Haxton} W.~C.,  1990,
  \mn@doi [\apj] {10.1086/168839}, \href
  {https://ui.adsabs.harvard.edu/abs/1990ApJ...356..272W} {356, 272}

\bibitem[\protect\citeauthoryear{{de Boer} et~al.,}{{de Boer}
  et~al.}{2012a}]{deBoer2012}
{de Boer} T.~J.~L.,  et~al., 2012a, \mn@doi [\aap]
  {10.1051/0004-6361/201118378}, \href
  {https://ui.adsabs.harvard.edu/abs/2012A&A...539A.103D} {539, A103}

\bibitem[\protect\citeauthoryear{{de Boer} et~al.,}{{de Boer}
  et~al.}{2012b}]{fornaxdeBoer2012}
{de Boer} T.~J.~L.,  et~al., 2012b, \mn@doi [\aap]
  {10.1051/0004-6361/201219547}, \href
  {https://ui.adsabs.harvard.edu/abs/2012A&A...544A..73D} {544, A73}

\bibitem[\protect\citeauthoryear{{de Boer}, {Belokurov}  \& {Koposov}}{{de
  Boer} et~al.}{2015}]{deBoer2015}
{de Boer} T.~J.~L.,  {Belokurov} V.,   {Koposov} S.,  2015, \mn@doi [\mnras]
  {10.1093/mnras/stv946}, \href
  {https://ui.adsabs.harvard.edu/abs/2015MNRAS.451.3489D} {451, 3489}

\makeatother
\end{thebibliography}

%\begin{thebibliography}{99}
%\end{thebibliography}

%%%%%%%%%%%%%%%%%%%%%%%%%%%%%%%%%%%%%%%%%%%%%%%%%%

%%%%%%%%%%%%%%%%% APPENDICES %%%%%%%%%%%%%%%%%

\bsp	% typesetting comment
\label{lastpage}
\end{document}